\pdfoutput=1

\RequirePackage{amsmath,amssymb,mathtools}

\documentclass[twocolumn,epjc3]{svjour3}

\usepackage{graphicx}
\usepackage{lineno}
\usepackage{xcolor}
\usepackage{xspace}
\usepackage{booktabs}
\usepackage{cite}
\usepackage[utf8]{inputenc}

\DeclareRobustCommand{\NNLOJET}{NNLO\scalebox{.8}{JET}\xspace}

\newcommand{\rd}{\ensuremath{\mathrm{d}}}

\def\l{\left}
\def\r{\right}
\def\be{\begin{equation}}
\def\ee{\end{equation}}
\def\ba{\begin{eqnarray}}
\def\ea{\end{eqnarray}}

\newcommand{\al}{\alpha}
\newcommand{\bt}{\beta}

\newcommand{\la}{\lambda}
\newcommand{\si}{\sigma}
\newcommand{\nn}{\nonumber}

\begin{document}

\renewcommand\makeheadbox{%
  \hfill
  CERN-TH-2019-140,
  IPPP/19/68,
  ZU-TH 39/19
}

\title{%
Second-order QCD corrections to event shape distributions in deep inelastic scattering}

\author{%
  T.~Gehrmann\thanksref{zurich}\and
  A.~Huss\thanksref{cern}\and
  J.~Mo\thanksref{zurich}\and
  J.~Niehues\thanksref{durham}
}

\raggedbottom

\institute{%
  Physik-Institut, Universit\"at Z\"urich, Winterthurerstrasse 190, CH-8057 Z\"urich, Switzerland \label{zurich}\and
  Theoretical Physics Department, CERN, CH-1211 Geneva 23, Switzerland \label{cern}\and
  Institute for Particle Physics Phenomenology, Durham University, Durham, DH1 3LE, United Kingdom \label{durham}}

\date{}

\maketitle

\begin{abstract}
We compute the next-to-next-to-leading order (NNLO) QCD corrections to event shape distributions
and their mean values in 
deep inelastic lepton-nucleon scattering. The magnitude and shape of the 
corrections varies considerably between different 
variables. The corrections reduce the renormalization and 
factorization scale uncertainty of the predictions. Using a 
dispersive model to describe non-perturbative power corrections, we 
compare the NNLO QCD predictions with data from the 
H1 and ZEUS experiments. The newly derived corrections 
improve the theory description of the
distributions and of their mean values.
\end{abstract}

\section{Introduction}
Event shape variables allow various kinematical properties of hadronic final states to be analysed. The resulting event shape distributions were measured extensively in $e^+e^-$~\cite{Kluth:2006bw} and $ep$~\cite{Newman:2013ada} collisions, enabling a variety of precision QCD studies, including measurements of the strong coupling constant, resummation and parton-shower effects, investigations of non-perturbative power corrections, and tuning of multi-purpose event simulation models. 

Precision studies of event shapes distributions demand that their  theoretical description is
of comparable accuracy to the experimental measurements, requiring the 
calculation of higher order contributions in perturbative QCD. For $e^+e^-$ event shapes,
an appropriate level of theory  precision was achieved already some time ago with 
the calculation of the next-to-next-to-leading order (NNLO) QCD corrections~\cite{GehrmannDeRidder:2007hr,GehrmannDeRidder:2009dp,Ridder:2014wza,Weinzierl:2009ms,Weinzierl:2009yz,DelDuca:2016ily,Gehrmann:2017xfb}
in the form of generic parton-level event generators that allow any infrared-safe event shape 
distribution to be computed. These fixed-order NNLO 
results can be combined with resummation of large logarithmic corrections to next-to-next-to-leading logarithmic 
level (NNLL) and beyond for specific event shape 
variables~\cite{Becher:2008cf,Chien:2010kc,Abbate:2010xh,Monni:2011gb,Becher:2012qc,Hoang:2014wka}.

For event shapes in deeply inelastic $ep$ scattering (DIS), the currently available level of theoretical accuracy is 
 lower, with fixed-order results only known to next-to-leading order (NLO)~\cite{Catani:1996vz,Graudenz:1997gv,Nagy:2001xb}
and resummation at next-to-leading logarithmic level (NLL)~\cite{Antonelli:1999kx,Dasgupta:2001eq,Dasgupta:2002dc,Dasgupta:2003iq}.
The theory uncertainty (as quantified through variation of the renormalization and factorization scales) 
on these predictions is often comparable to or larger than the experimental errors on the event shape measurements from 
H1~\cite{Aktas:2005tz} and ZEUS~\cite{Chekanov:2006hv}, thereby  limiting the extraction of fundamental QCD parameters 
from these data. To overcome this limitation requires an improvement of the fixed-order 
predictions to NNLO, which is presented in the following. 

This paper is structured as follows. In Section~\ref{sec:shapes}, we summarise the definitions of the most common DIS 
event shape variables, and the kinematical ranges covered by the 
 H1~\cite{Aktas:2005tz} and ZEUS~\cite{Chekanov:2006hv} measurements. The calculation of NNLO
 corrections to event shape distributions is performed in the \NNLOJET 
framework~\cite{Gehrmann:2018szu} and follows closely the related NNLO calculations of jet production in 
DIS~\cite{Currie:2016ytq,Currie:2017tpe} and is documented in Section~\ref{sec:calc}. 
To compare the resulting parton-level NNLO 
predictions with experimental hadron-level data, we employ a dispersive model~\cite{Dokshitzer:1995qm,Dasgupta:1997ex,Dasgupta:1998xt},
described in Section~\ref{sec:power}, determining the non-perturbative power corrections to the event shape distributions. We perform detailed comparisons of the hadron-level predictions to event shape data from H1 and ZEUS in Section~\ref{sec:results}. Our findings are summarized in Section~\ref{sec:conc}.

\section{Event shape variables}
\label{sec:shapes}
Event shapes in DIS are measured in the Breit frame, defined by the momentum directions of the virtual photon (current axis) and the proton (remnant axis), and 
boosted such that the energy component of the virtual photon momentum vanishes. The Breit frame provides a separation in pseudorapidity
$\eta$ between the proton remnant (remnant hemisphere, $\eta>0$) and the hard scattering process (current hemisphere, $\eta<0$). 
The event shape variables are dimensionless quantities that are determined from the four-momenta  $p_h= (E_h,\vec{p}_h)$ of all particles  in the current hemisphere. 
The different variables~\cite{Dasgupta:2003iq}, which are generically denoted as $F$, are defined as follows.

The thrust $\tau_\gamma$ measures the longitudinal momentum components projected onto the current axis:
\be
\tau_\gamma = 1-T_\gamma, \quad \mathrm{with} \quad T_\gamma = \frac{\sum_h |{p}_{z,h}|}{\sum_h |\vec{p}_h|}\,.
\ee

Thrust $\tau_T$ is the thrust with respect to the thrust axis in the direction $\vec{n}_T$ which maximizes the longitudinal momentum components projected onto this axis:
\be
\tau_T = 1 - T_T, \quad \mathrm{with} \quad T_T = \max_{\vec{n}_T}\frac{\sum_h |\vec{p}_h \cdot \vec{n}_T|}{\sum_h |\vec{p}_h|}\,.
\ee
This is analogous to the  definition of thrust in $e^+ e^-$ collisions. 

The jet mass parameter $\rho$ is the squared invariant mass in the current hemisphere, normalized to  
four times the total energy squared:
\be
\rho = \frac{(\sum_h p_h)^2}{(2 \sum_h E_h)^2}\,.
\ee

The jet broadening $B_\gamma$ measures the sum of the transverse momenta with respect to the current axis:
\be
B_\gamma = \frac{\sum_h |\vec{p}_{t,h}|}{2 \sum_h |\vec{p}_h|}\,.
\ee

As with thrust, the jet broadening can also be defined with respect to the thrust axis:
\be
B_T = \frac{\sum_h |\vec{p}_h \times \vec{n}_T|}{2\sum_h |\vec{p}_h|}\,.
\ee

Finally, the $C$-parameter is derived from the linear momentum tensor $\Theta^{ij}$:
\be
\Theta^{ij} = \frac{1}{\sum_h |\vec{p}_h|} \sum_h\frac{ p_h^i p_h^j}{|\vec{p}_h|}\,.
\ee
with the  eigenvalues $\la_1, \la_2, \la_3$ of $\Theta^{ij}$ yielding
\be
C = 3(\la_1 \la_2 + \la_2 \la_3 + \la_3 \la_1)\,.
\ee
Equivalently, it can be expressed as
\be
C = \frac{3}{2} \frac{\sum_{h,h'} |\vec{p}_h| |\vec{p}_{h'}| \sin ^2 \theta_{hh'}}{(\sum_h |\vec{p}_h|)^2}\,,
\ee
where $\theta_{hh'}$ is the angle between particles $h$ and $h'$.

In the experimental analysis, the event shapes are computed from the hadron momenta in the current hemisphere, while the theoretical calculation uses the parton momenta. 
For the Born-level contribution to inclusive DIS, lepton-quark scattering, only the final state quark is produced in the current hemisphere, with thrust axis and current axis coinciding. Consequently, 
all event shape variables defined above become trivially zero. The first non-trivial contribution to the event shape distributions arises from two-parton final states: $eq \to eqg$ or $eg \to eq\bar q$, such 
that the   leading-order (LO) perturbative contribution is ${\cal O}(\alpha_s)$. The event shape distributions are thus closely related to DIS two-jet production in the Breit frame. 

In higher-multiplicity final states, it is possible that all partons scatter into the remnant hemisphere, leaving the current hemisphere empty. To ensure infrared safety of the observables, these events are not accepted by demanding that the total energy in the current hemisphere of an event exceeds some minimum value $\epsilon_{\mathrm{lim}}$
\be
\sum_h E_h > \epsilon_{\mathrm{lim}}\,.
\ee

Event shapes in deep inelastic scattering have been measured at HERA by the  H1~\cite{Aktas:2005tz} and ZEUS~\cite{Chekanov:2006hv} experiments, based on the analysis of
electron-proton scattering data taken at a centre-of-mass energy of $\sqrt{s}=319$~GeV (the H1 data set also contains a small fraction of data taken at $\sqrt{s} = 301$~GeV). 
The DIS kinematics in the process $e(k) + p(p) \to e(k') + X (p_X)$, with momentum transfer $q=k'-k$ is described by the variables $Q^2=-q^2$, $x=Q^2/(2q\cdot p)$ and $y=Q^2/(x s)$.

The H1 analysis~\cite{Aktas:2005tz} selects events with
\begin{equation}
0.1<y<0.7\\, \quad 196~\mbox{GeV}^2 < Q^2 < 40000~\mbox{GeV}^2\,,
\label{eq:h1cuts}
\end{equation}
which are then classified into bins in $Q=\sqrt{Q^2}$, as listed in Table~\ref{tab:H1KinRange}.  For the event shape determination, 
$\epsilon_{\mathrm{lim}}= {Q}/{10}$ is used. 
\begin{table}
	\centering
	\begin{tabular}{ c  r@{\,--\,}l }
		\toprule
		Bin & \multicolumn{2}{c}{$Q (\mathrm{GeV})$}  \\ 
		\midrule
		1  & 14  & 16   \\
		2  & 16  & 20   \\
		3  & 20  & 30   \\
		4  & 30  & 50   \\
		5  & 50  & 70   \\
		6  & 70  & 100  \\
		7  & 100 & 200  \\
		\bottomrule
	\end{tabular}
	\caption{Kinematic boundaries of the bins in $Q^2$ in the H1 analysis~\protect{\cite{Aktas:2005tz}}.}
	\label{tab:H1KinRange}
\end{table}
\begin{table}
	\centering
	\begin{tabular}{ c r@{\,--\,}l r@{\,--\,}l }
		\toprule
		Bin & \multicolumn{2}{c}{$Q (\mathrm{GeV})$} & \multicolumn{2}{c}{$x$} \\ 
		\midrule
		1  & 80    & 160    &  0.0024 & 0.010  \\
		2  & 160   & 320    &  0.0024 & 0.010  \\
		3  & 320   & 640    &  0.01   & 0.05   \\
		4  & 640   & 1280   &  0.01   & 0.05   \\
		5  & 1280  & 2560   &  0.025  & 0.150  \\
		6  & 2560  & 5120   &  0.05   & 0.25   \\
		7  & 5120  & 10240  &  0.06   & 0.40   \\
		8  & 10240 & 20480  &  0.10   & 0.60   \\
		\bottomrule
	\end{tabular}
	\caption{Kinematic boundaries of the bins in $Q^2$ and $x$ in the ZEUS analysis~\protect{\cite{Chekanov:2006hv}}.}
	\label{tab:ZeusKinRange}
\end{table}

The ZEUS analysis~\cite{Chekanov:2006hv} covers the kinematic range
\begin{eqnarray}
0.0024 < x < 0.6\,, \quad 0.04 < y < 0.9\,, \nonumber\\
80~\mbox{GeV}^2 < Q^2 < 20480~\mbox{GeV}^2\,,
\label{eq:zeuscuts}
\end{eqnarray}
with events binned into in $(Q^2,x)$, described in Table~\ref{tab:ZeusKinRange}. The energy cut in the current hemisphere used by ZEUS is $\epsilon_{\mathrm{lim}}= {Q}/{4}$. 

Both experiments normalize the event shape distributions to the DIS cross section integrated over the kinematical bin under consideration, which is determined without applying the $\epsilon_{\mathrm{lim}}$ cut. 

Both experiments performed measurements~\cite{Aktas:2005tz,Chekanov:2006hv} of the event shape distributions for $F=\tau_\gamma$, $\tau_T$, $\rho$, $B_\gamma$, $C$. In addition, they also measured the mean values  $\langle F\rangle$ 
for these variables, supplemented in the ZEUS  study  by a measurement of the mean value $\langle B_T \rangle$ of the jet broadening with respect to the thrust axis. The measurements 
of the mean values are done for the same kinematical bins, Tables~\ref{tab:H1KinRange}--\ref{tab:ZeusKinRange}, as used for the distributions.

\section{QCD corrections to event shapes}
\label{sec:calc}
The event shape variables defined above assume non-trivial values only for final states containing two or more partons. Consequently, the event shape 
distributions in DIS receive the same parton-level contributions as two-jet production in DIS. Higher-order QCD corrections to event shape distributions can thus 
be obtained from the corresponding calculation for di-jet production by replacing the jet reconstruction algorithm by computations of the event shape variables.

We calculate the differential distributions and mean values for the DIS event shapes with the parton-level Monte Carlo event generator \NNLOJET, by extending the existing calculation of 
NNLO corrections to di-jet production in DIS~\cite{Currie:2016ytq,Currie:2017tpe}. It combines the contributions from four-parton production at tree-level~\cite{Hagiwara:1988pp,Berends:1988yn,Falck:1989uz}, 
three-parton production at one loop~\cite{Glover:1996eh,Bern:1996ka,Campbell:1997tv,Bern:1997sc} and two-parton production at
 two loops~\cite{Garland:2001tf,Garland:2002ak,Gehrmann:2002zr,Gehrmann:2009vu}, using the antenna subtraction method~\cite{GehrmannDeRidder:2005cm,Daleo:2006xa,Currie:2013vh} 
 to isolate infrared singular terms from the 
different contributions, which are then combined to yield numerically finite predictions for arbitrary infrared-safe observables constructed from the parton momenta. 
Besides for di-jet production at NNLO, the same ingredients and setup have been used previously in the computation of N$^3$LO corrections to single jet production in DIS~\cite{Currie:2018fgr}, in 
extractions of the strong coupling constant from DIS jet data~\cite{Andreev:2017vxu,Britzger:2019kkb}, and in studies of diffractive di-jet production~\cite{Britzger:2018zvv}. 
The calculations have also been extended to 
jet production in charged current DIS~\cite{Niehues:2018was,Gehrmann:2018odt} at the same perturbative orders.

We compute the event shapes for electron-proton collisions with $\sqrt{s}=319$~GeV, using the NNPDF3.1 parton distributions with $\alpha_s(M_Z)=0.118$ and for $N_F=5$ massless quark flavours.
 Central renormalization 
and factorization scales are fixed to $\mu_F=\mu_R=Q$, and theory uncertainties are estimated by the envelope of 
varying these scales independently by a factor two up and down, avoiding the pairings of 
variations in opposite directions (seven-point scale variation). Event selection cuts on the lepton variables and on $\sum_h E_h$ are applied according to the 
H1~\cite{Aktas:2005tz} and ZEUS~\cite{Chekanov:2006hv}  analyses, and events are then classified into the different kinematical bins of Tables~\ref{tab:H1KinRange} and~\ref{tab:ZeusKinRange}.
The total hadronic DIS
cross section for each kinematical bin (required for the normalization of the event shape distributions and mean values) 
is obtained to NNLO from \NNLOJET, based on the one-jet calculation to this order~\cite{Currie:2018fgr}. Central renormalization and factorization scales are used for the normalization.

\subsection{Event shape distributions}

The event shape distributions are computed as histograms in the event shape variables. We use a considerably finer bin resolution than in the experimental analyses~\cite{Aktas:2005tz,Chekanov:2006hv}, 
which will subsequently allow us to apply hadronization corrections that result in a dynamical shift of the event shape variables. The histograms are defined in terms of variable ranges and number of 
equal-sized bins:
\ba
\tau_\gamma:& [0,1], &100\,, \nn\\
\tau_T:& [0,0.5], &100\,, \nn\\
\rho:& [0,0.25], &80(\mbox{H1})/100(\mbox{ZEUS})\,, \nn\\
B_\gamma:& [0,0.5], &100 \,, \nn\\
C:& [0,1], &100 \,.
\label{eq:hires}
\ea

The fixed-order calculation for an event shape $F$ diverges in the limit $F\to 0$, 
where all-orders resummation of large $\log(F)$-terms is required. In this limit, the fixed-order expressions become meaningless, and we 
accordingly apply 
cuts on the minimum values of each shape variable, which set the first few bins of the distributions to zero:  
\ba
\tau_\gamma &\ge \tau_\gamma^{\mathrm{cut}} &= 0.05\,, \nn\\
\tau_T &\ge \tau_T^{\mathrm{cut}} &= 0.025\,, \nn\\
\rho &\ge \rho^{\mathrm{cut}} &= 0.01\,, \nn\\
B_\gamma &\ge B_\gamma^{\mathrm{cut}} &= 0.05 \,,\nn\\
C &\ge C^{\mathrm{cut}} &= 0.05\,.
\label{eq:ircuts}
\ea
These cuts are typically within the first bin of the experimental analysis, which should anyhow be discarded in the comparison of fixed-order theory and experimental data. 
\begin{figure}[t]
 \hspace{-0.2in} \includegraphics[width=3.8in]{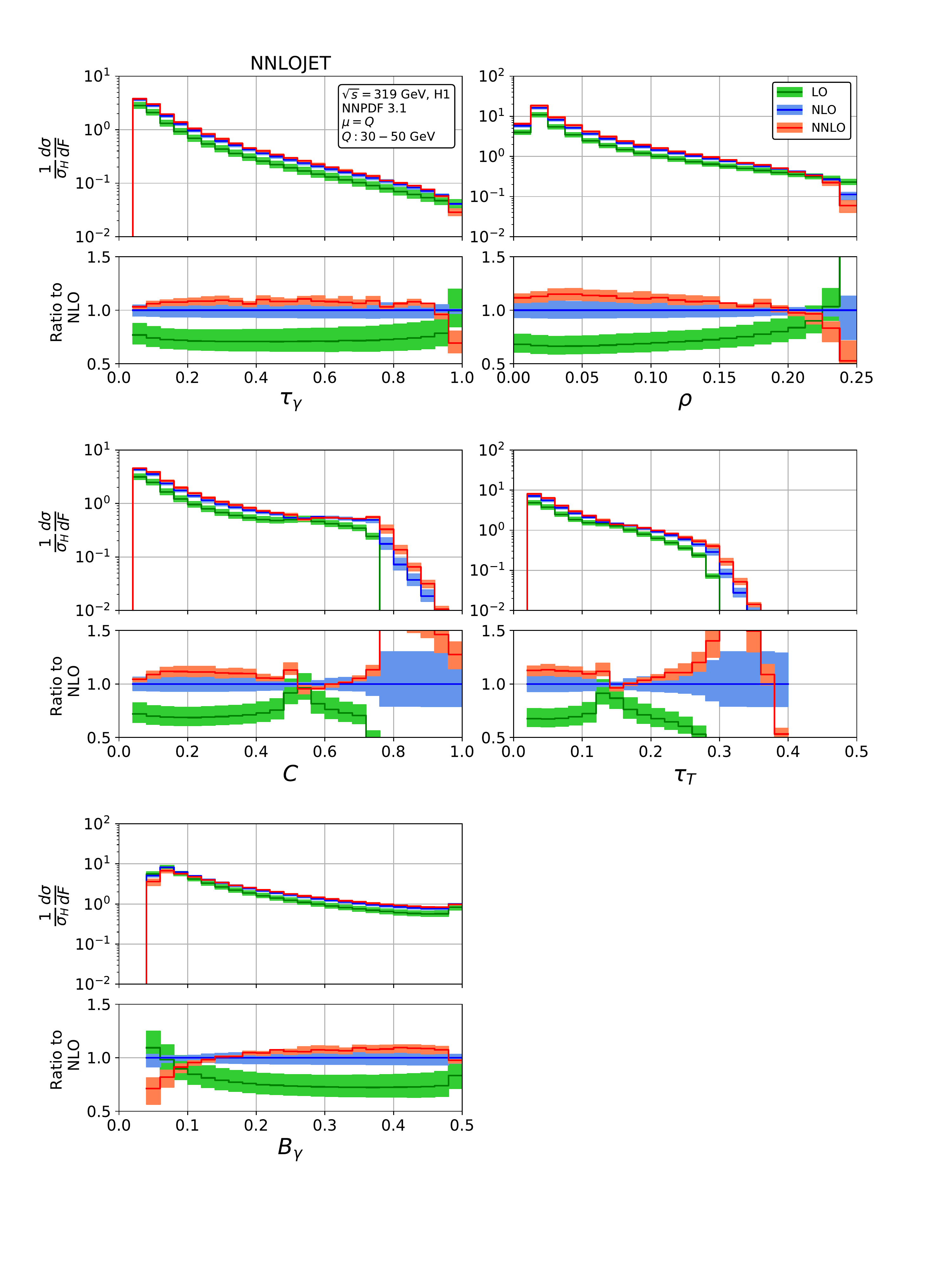}
  \caption{Fixed-order predictions for the event shape distribution for H1 kinematics~\protect{\cite{Aktas:2005tz}} in $Q=30-50$~GeV bin:
 LO (green), NLO (blue) and NNLO (red), for H1 kinematics~\protect{\cite{Aktas:2005tz}}. The lower 
  frames display the ratio to the NLO predictions for the central scale $\mu^2=Q^2$.}
  \label{fig:H1FO}
\end{figure}

Figure~\ref{fig:H1FO} displays the fixed-order predictions (re-binned from the initial histograms by combining four adjacent bins each) for the H1 kinematics in the 
 $Q=30-50$~GeV bin. 
Since the qualitative behaviour of the higher-order corrections to the distributions is similar for all kinematical 
bins, we show the fixed-order distributions without power corrections only for one representative bin. 
 The quantitative size of the corrections and of their uncertainties decreases with increasing $Q$, mainly due to the decrease in the running coupling constant $\alpha_s$. 
 
In general, we observe that the NNLO corrections in the bulk of all distributions are typically positive (up to +20\%), often displaying only a marginal or no 
overlap of the uncertainty bands at NLO and NNLO. The scale 
uncertainty decreases from NLO ($\sim$10\%) to NNLO ($\sim$5\%).
Even in the bulk, the higher-order corrections are not uniform between the distributions, each displaying a non-trivial shape in the NNLO/NLO ratio.

Towards the kinematical edges $F\to 0$ and $F\to F_{{\rm max}}$,  the higher-order corrections behave differently for  each distribution, often displaying large effects well beyond the 
scale uncertainty estimates. For $F\to F_{{\rm max}}$, these features are caused by two different but related issues. For some of the shape variables, $F_{{\rm max}}$ can not yet be realised in the 
Born process, owing to its low multiplicity. This is the case for the $C$-parameter which has a Born-level upper limit of $3/4$ and for  $\tau_T$ with an upper limit of $0.293$. Higher order real radiation 
corrections allow to attain larger values of $F$, thereby resulting in a kinematical mismatch between real and virtual contributions (Sudakov shoulder,~\cite{Catani:1997xc}),
 which (although finite) produces large 
perturbative corrections in the vicinity of the Born-level kinematical limit. 

In the case of DIS event shape variables, the kinematical constraints of the Born process produce further structures that narrow down the 
dimensionality of the final state phase space for specific values of different variables. These ridges in the multi-dimensional phase space were investigated in detail in~\cite{Dasgupta:2002dc} and 
produce kinks and spikes in the one-dimensional event shape distributions. These are sometimes already present at leading order, and go along with large and unstable higher order corrections in the immediate 
vicinity of the exceptional points, which are visible in particular in the distributions in $C$ and $\tau_T$ in Figure~\ref{fig:H1FO}.
These features are typically localised in small patches of the phase space. For sufficiently large bin sizes, their impact is diluted to an invisible level. High-resolution measurements 
of event shape distributions, for example at a future electron-ion collider~\cite{Accardi:2012qut} or at the LHeC~\cite{AbelleiraFernandez:2012cc}
will be able to resolve these features, thereby potentially necessitating resummation of large corrections associated with them. 

At low values of $F$, the fixed-order predictions contain logarithmic terms $\log F$ at each order in perturbation theory, which spoil the convergence of the fixed-order perturbative expansion. In Figure~\ref{fig:H1FO}, the onset of these effects is visible in particular in the $B_\gamma$ distribution, while its onset takes place only at lower values of $F$ in all other distributions. A description of the event shape distributions over the full kinematical range, and extending towards lower values of $F$ than probed by currently available measurements~\cite{Aktas:2005tz,Chekanov:2006hv} will need to 
include the resummation of these $\log F$ terms, which is currently known to next-to-leading logarithmic level~\cite{Antonelli:1999kx,Dasgupta:2001eq,Dasgupta:2002dc,Dasgupta:2003iq} for all distributions. 

\subsection{Mean values}
\label{sec:resultmean}
\begin{figure}[t]
	\centering
	\includegraphics[width=3.5in]{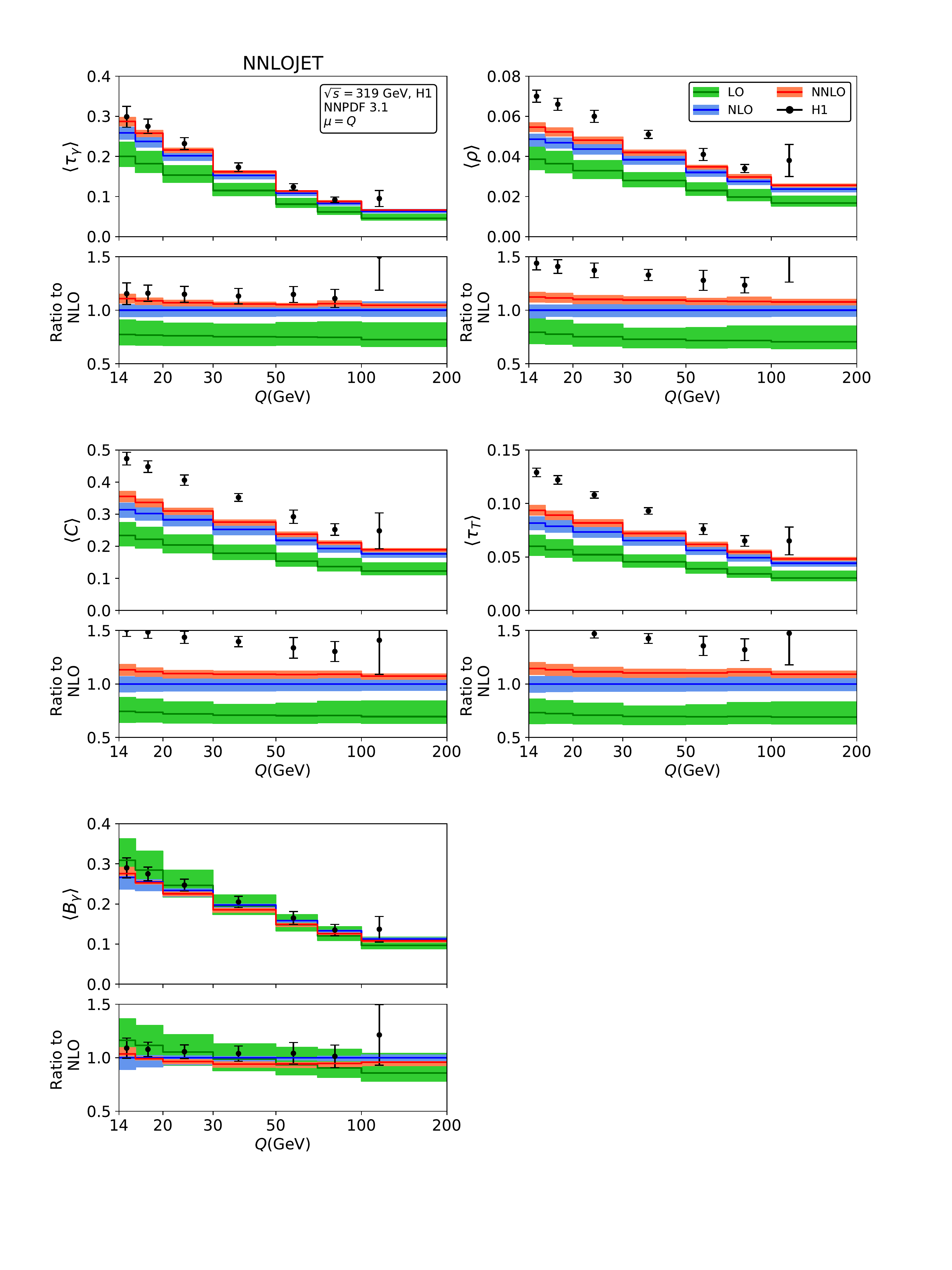}
	\caption{Fixed-order predictions for the mean value of the event shapes at LO (green), NLO (blue) and NNLO (red) compared to  H1 data~\protect\cite{Aktas:2005tz}. }
		\label{fig:H1meanFO}
\end{figure}
\begin{figure}[t]
	\centering
	\includegraphics[width=3.5in]{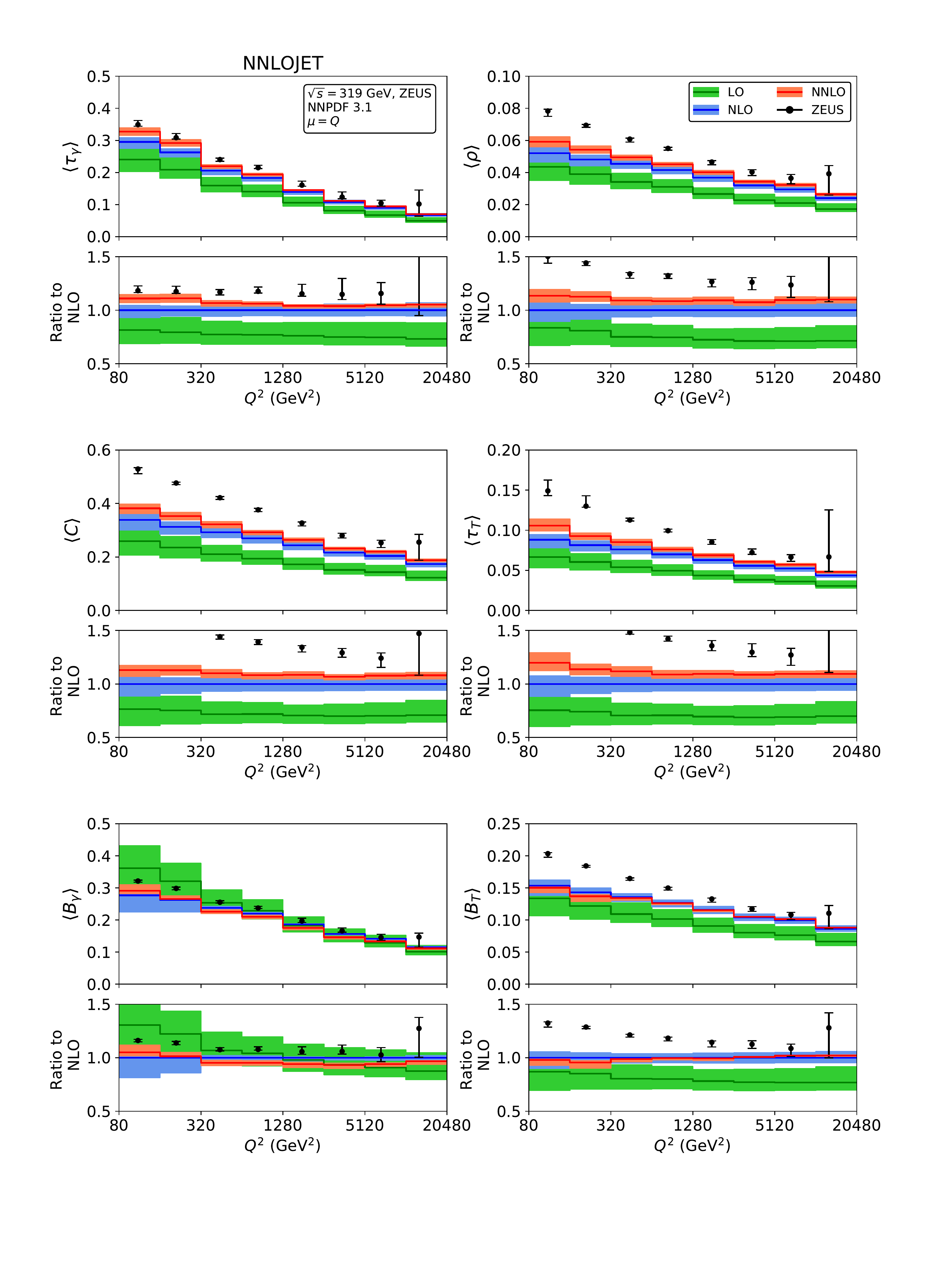}
	\caption{Fixed-order predictions for the mean value of the event shapes at LO (green), NLO (blue) and NNLO (red) compared to ZEUS data~\protect\cite{Chekanov:2006hv}. } 
		\label{fig:ZEUSmeanFO}
\end{figure}

The mean values of the different event shapes variables are computed using  \NNLOJET by weighting each event with the reconstructed value of the event shape variable under consideration. 
The phase space integrations are performed by imposing only a very low technical cut-off of $F_{{\rm min}}=0.001$ 
on the event shape variables, since the weighting with the shape variable regulates the divergent behaviour of the integrals for 
$F\to 0$, rendering the mean value integrals finite. The mean values are also normalized to the inclusive hadronic cross sections. 

The fixed-order predictions for the mean values are displayed in Figures~\ref{fig:H1meanFO}--\ref{fig:ZEUSmeanFO}, for the H1~\cite{Aktas:2005tz}and ZEUS~\cite{Chekanov:2006hv} kinematics. 
With the exception of the broadenings $\langle B_\gamma\rangle$ and $\langle B_T\rangle$, the NNLO corrections to the mean values are positive for all event shapes, and decrease in 
magnitude with increasing $Q^2$. The NNLO predictions are often at the upper boundary of the NLO theory uncertainty band, for the lowest $Q^2$ bins they are even outside the NLO band. For the broadenings, the NNLO corrections to   $\langle B_\gamma\rangle$ are positive at low $Q^2$, and become negative at large $Q^2$, to $\langle B_T\rangle$
they display the opposite behaviour, and are smaller in absolute magnitude. For all mean values, inclusion of the NNLO corrections leads to a reduction of the scale uncertainty compared to NLO, which 
is most pronounced for the broadenings, whereas being more modest for the other shape variables. For large values of $Q^2>2500$~GeV$^2$, the NNLO theory uncertainty is limited to below 5\%. 

Comparing the fixed-order predictions to the measurements of the mean values from H1 and ZEUS, we observe that the data are considerably above the theory predictions throughout all shape variables and 
for all values of $Q^2$, although the discrepancy is most pronounced at low $Q^2$. This behaviour indicates the relevance of power corrections from hadronization effects, which can have large effects 
on the mean values~\cite{Dokshitzer:1995qm,Dasgupta:1997ex,Dasgupta:1998xt}. 
\section{Hadronization effects}
\label{sec:power}

In the previous section, we computed higher-order corrections to the DIS event shape distributions and mean values at parton level. To compare these predictions with hadron-level data requires accounting for
 the impact of the parton-hadron transition, which is a non-perturbative process. Consequently, these hadronization effects cannot be computed in perturbation theory, but require a non-perturbative model description. 
The hadronization corrections are expected to be suppressed by positive powers of $\Lambda/Q$, such that their relative numerical impact is decreasing with increasing $Q$. In the following, we employ  the dispersive model~\cite{Dokshitzer:1995qm} to estimate the leading
 power corrections at order $(\Lambda/Q)$  to event shape distributions. This model has been worked out in detail for the DIS event shapes in Ref.~\cite{Dasgupta:1997ex}, and its 
implications are 
briefly summarized in the following. 

In the dispersive model, an effective coupling $\alpha_{\mathrm{eff}}$ is introduced at low scales, which is matched to the running QCD coupling $\alpha_s(\mu)$ at a scale $\mu_I = 2$ GeV. This gives a constant $\alpha_0$ which is defined as the first moment of the effective coupling below the scale $\mu_I$,
\be
\alpha_0(\mu_I) = \frac{1}{\mu_I} \int_0^{\mu_I} \rd\mu \, \alpha_{\mathrm{eff}}(\mu)\,.
\ee
The power corrections are suppressed by powers of $1/Q$, and result in a shift of the perturbative differential distribution
\be
\frac{\rd\si^{\mathrm{hadron}}(F)}{\rd F} = \frac{\rd\si^{\mathrm{parton}}(F - a_F P)}{\rd F},
\label{eq:powershift}
\ee
where the power correction $P$ is universal for all the event shape variables. The perturbative ingredients to the dispersive model are the running of the coupling constant and the relation between the 
$\overline{{\rm MS}}$-coupling and the effective coupling, whose definition~\cite{Catani:1990rr}  absorbs universal correction terms from the cusp anomalous dimension. It can be expanded in 
$\al_s(Q)$, and its expression up to NNLO is given by~\cite{Davison:2008vx,Gehrmann:2009eh}: 
\ba
P &=&\frac{8 C_F}{\pi^2} \mathcal{M} \frac{\mu_I}{Q} \bigg\{ \al_0(\mu_I) - \al_s(Q) \nonumber \\ &&- \frac{\bt_0}{2\pi} \l( \log  \frac{Q}{\mu_I} + \frac{K}{\bt_0} + 1 \r) \al_s^2(Q) \nonumber \\ &&
- \bigg[ \frac{\bt_1}{2} \l( \log \frac{Q}{\mu_I} + \frac{2 L}{\bt_1} + 1 \r) \nonumber \\ &&
+ 2\bt_0^2 \l( \log  \frac{Q}{\mu_I} + \frac{K}{\bt_0} + 1 \r) \nonumber \\ &&
+ \bt_0^2 \log \frac{Q}{\mu_I} \l( \log \frac{Q}{\mu_I} + \frac{2 K}{\bt_0} \r) \bigg] \frac{\al_s^3(Q)}{4\pi^2} \bigg\},
\label{eq:powerP}
\ea
with $\mathcal{M} = 1.49 $ a constant normalization factor (Milan factor~\cite{Dokshitzer:1998pt,Dasgupta:1998xt}) accounting for 
higher-order contributions. In our numerical results, we use $\al_0(\mu_I) = 0.5$ at $\mu_I=2$~GeV, which 
has been estimated from fits to event shape moments in DIS~\cite{Adloff:1999gn,Aktas:2005tz,Chekanov:2006hv} and 
$e^+e^-$ annihilation~\cite{Gehrmann:2009eh}. 
The beta-function coefficient  and cusp anomalous dimension~\cite{Moch:2004pa} in the above expression are:
\ba
\bt_0 &=& \frac{11}{3} C_A - \frac{4}{3} T_F N_F, \nonumber \\ 
\bt_1 &=& \frac{34}{3} C_A^2 - \frac{20}{3} C_A T_F N_F - 4 C_F T_F N_F, \nonumber \\ 
K &=& \left(\frac{67}{18} - \frac{\pi^2}{6}\right)C_A - \frac{10 T_F N_F}{9}, \nonumber \\ 
L &=& C_A^2 \l(\frac{245}{24} - \frac{67}{9} \frac{\pi^2}{6} + \frac{11}{6} \zeta_3 + \frac{11}{5} \bigg( \frac{\pi^2}{6} \bigg)^2 \r) \nonumber \\ 
&+& C_A N_F \l( -\frac{209}{108} + \frac{10}{9} \frac{\pi^2}{6} - \frac{7}{3} \zeta_3 \r) \nonumber \\
&+& C_F N_F \l( -\frac{55}{24} + 2 \zeta_3 \r) + N_F^2 \l( -\frac{1}{27} \r),
\ea 
with $C_A=3$, $C_F=4/3$, $T_F=1/2$. 
The coefficients $a_F$ depend on the event shape variable, they were computed in~\cite{Dasgupta:1997ex} and are tabulated in~\cite{Adloff:1999gn}. Their values are repeated here:
\ba
a_{\tau_\gamma} &=& 1\,, \quad  
a_{\tau_T} = 1\,, \quad  
a_\rho = \frac{1}{2}\,, \nn\\
a_{B_\gamma} &=& \frac{1}{2} a'_B\,, \quad 
a_{B_T} = \frac{1}{2} a'_B\,, \quad
a_C = \frac{3}{2} \pi\,,
\label{eq:apower}
\ea
where the shift of the jet broadening has an additional enhancement~\cite{Dokshitzer:1998qp} given by
\be
a'_B = \frac{\pi}{2 \sqrt{2C_F \al_s (1+\frac{K}{2\pi} \al_s)}} + \frac{3}{4} - \frac{\bt_0}{12C_F} + \eta_0,
\ee
with $\al_s$ evaluated at the scale $\mu=e^{-\frac{3}{4}} \mu_R$ and $\eta_0 = -0.614$. For this analysis $a'_B$ varies from $1.6-2.3$. 

The dispersive model  is based on an analytic treatment of hadronization effects on a two-parton correlator~\cite{Dokshitzer:1995qm}, which corresponds to the mean value integral 
of each event shape. The effect of the dispersive 
power correction on the mean values is additive:
\be
\langle F \rangle = \langle F \rangle^{{\rm pert.}} + a_F\,  P,
\label{eq:powermean}
\ee
where $\langle F \rangle^{{\rm pert.}}$ is the mean value obtained in fixed-order perturbation theory, described in Section~\ref{sec:resultmean} above.
\begin{figure}[t]
\hspace{-0.2in}  \includegraphics[width=3.8in,page=1]{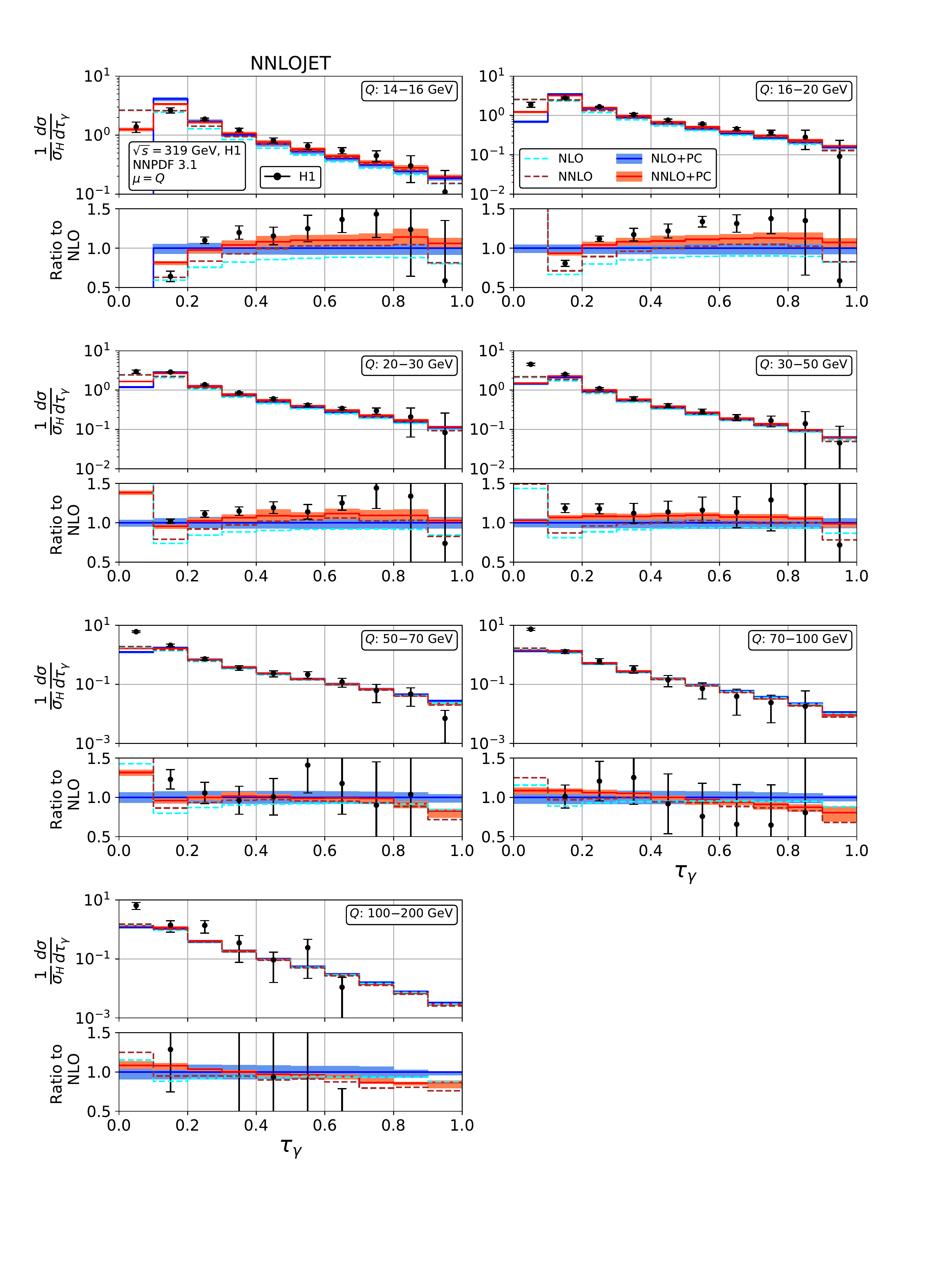}
  \caption{Event shape distribution for thrust with respect to boson axis: $\tau_\gamma$ fixed-order predictions at NLO (dashed cyan), NNLO (dashed brown), and corrected 
  for hadronization effects at NLO (blue) and NNLO (red), compared to H1 data~\protect{\cite{Aktas:2005tz}}. The lower 
  frames display the ratio to the NLO prediction for the central scale $\mu^2=Q^2$.}
  \label{fig:H1TBP}
\end{figure}
\begin{figure}[t]
 \hspace{-0.2in}  \includegraphics[width=3.8in,page=1]{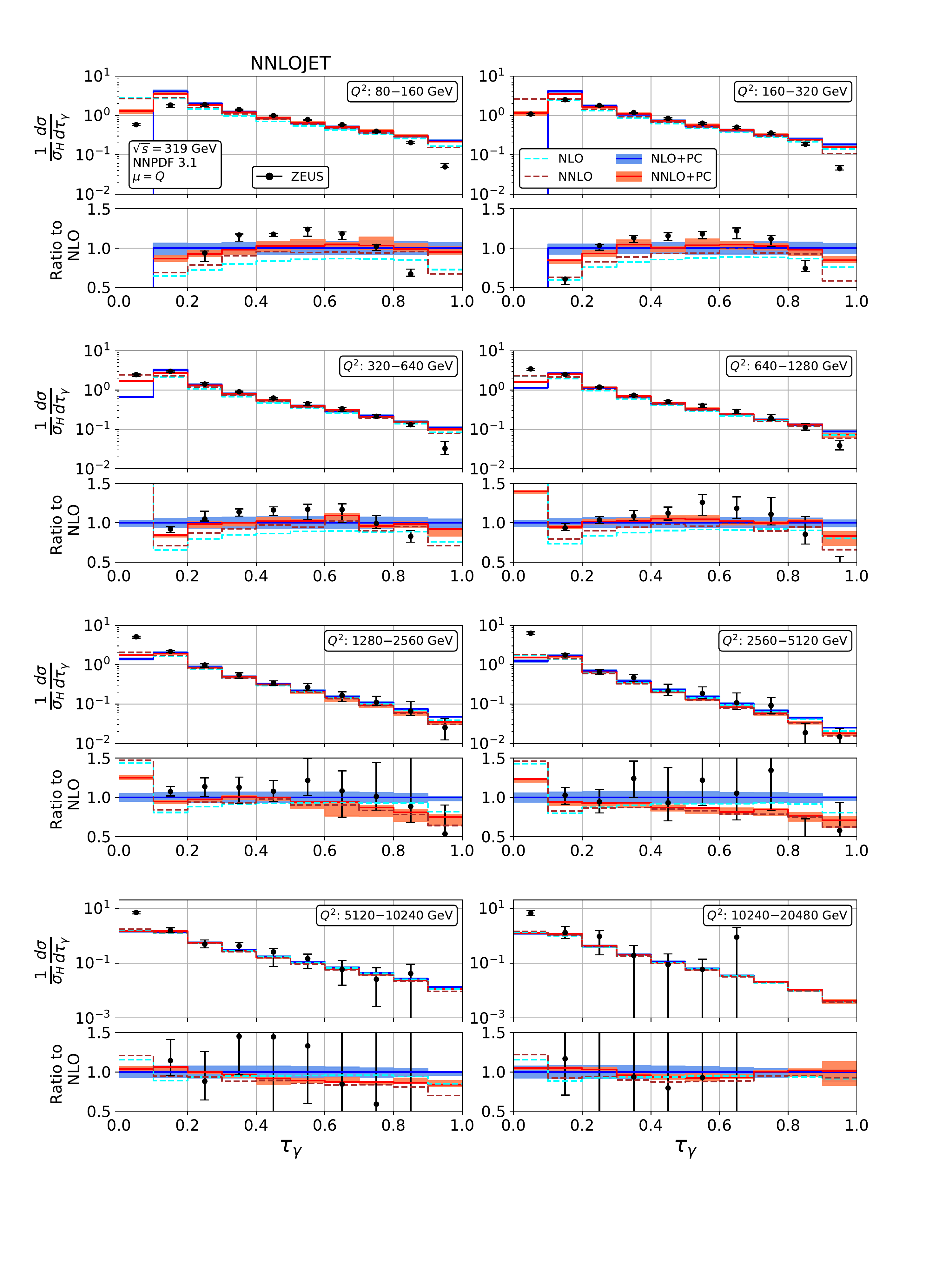}
  \caption{Event shape distribution for thrust with respect to boson axis: $\tau_\gamma$ fixed-order predictions at NLO (dashed cyan) and NNLO (dashed brown), and corrected 
  for hadronization effects at NLO (blue) and NNLO (red), compared to ZEUS data~\protect{\cite{Chekanov:2006hv}}. The lower 
  frames display the ratio to the NLO prediction for the central scale $\mu^2=Q^2$.}
  \label{fig:ZEUSTBP}
\end{figure}

 When applied to differential event shape distributions, 
the power correction $P$ in the shift~(\ref{eq:powershift}) can in principle depend on the numerical value of $F$. Using a constant shift $P$ for the full distribution amounts to an approximation, which 
may be overcome by an  improved treatment of the non-perturbative corrections. 

In combining the fixed-order predictions derived in the previous section with the power corrections, we truncate the factor $P$ in (\ref{eq:powerP})
to $\alpha_s^2(Q)$ for NLO, and to $\alpha_s^3(Q)$ for NNLO. Inclusion of the $\alpha_s^3(Q)$ terms leads to a substantial reduction of $P$, with $P_{{\rm NNLO}} \approx 0.60 \, P_{{\rm NLO}} $
at $Q=15$~GeV and  $P_{{\rm NNLO}} \approx 0.75 \, P_{{\rm NLO}} $ at $Q=100$~GeV.

\section{Results}
\label{sec:results}
With the inclusion of power corrections, the fixed-order parton-level predictions can now be compared to hadron-level data from the HERA 
experiments on event shape distributions and mean values. 

\subsection{Event shape distributions}

\begin{figure}[t]
\hspace{-0.2in}   \includegraphics[width=3.8in,page=3]{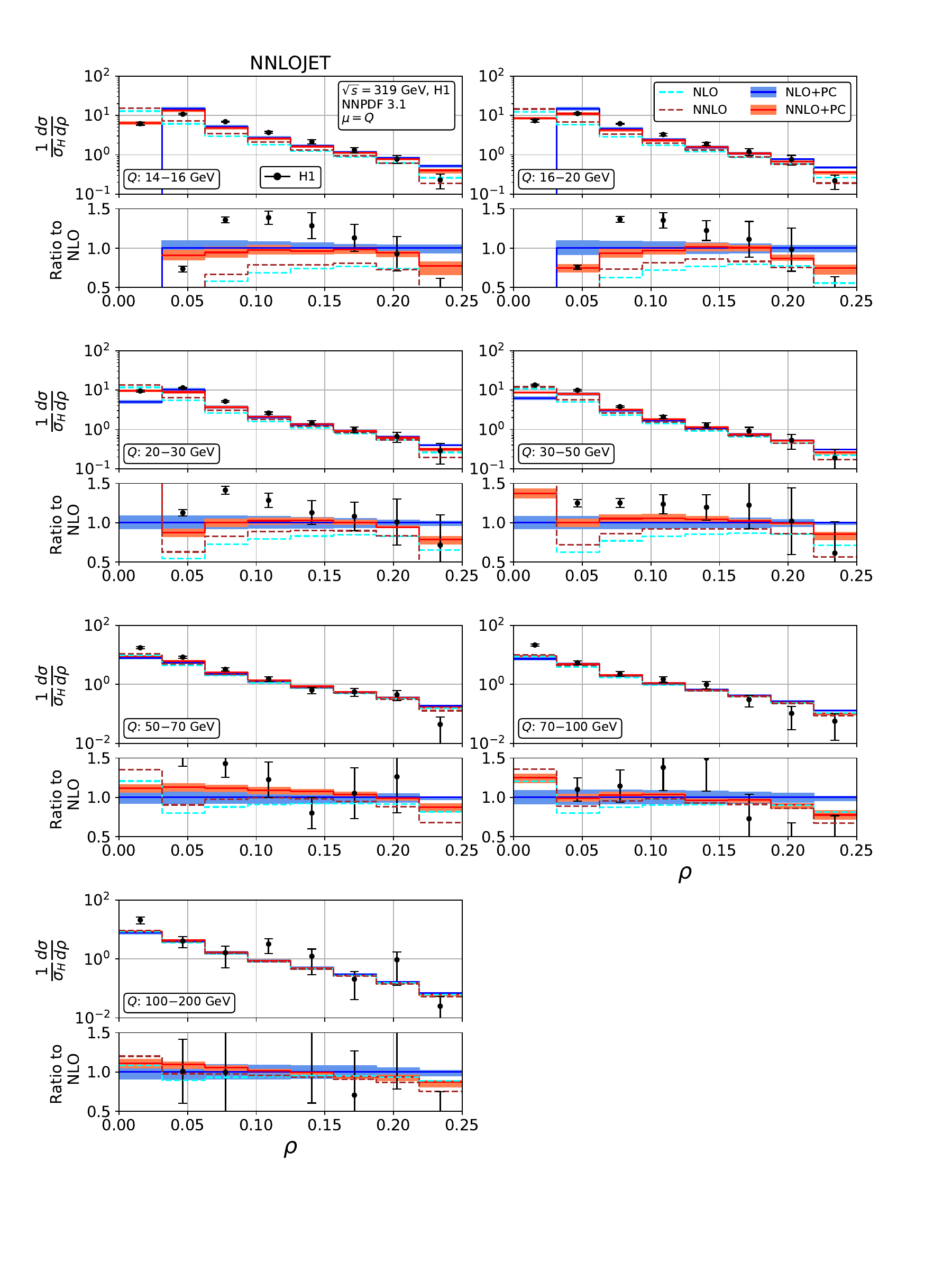}
  \caption{Event shape distribution for jet mass: $\rho$  fixed-order predictions at NLO (dashed cyan), NNLO (dashed brown), and corrected 
  	for hadronization effects at NLO (blue) and NNLO (red), compared to H1 data~\protect{\cite{Aktas:2005tz}}. The lower 
  frames display the ratio to the NLO prediction for the central scale $\mu^2=Q^2$.} 
  \label{fig:H1TMP}
\end{figure}
\begin{figure}[t]
\hspace{-0.2in}   \includegraphics[width=3.8in,page=3]{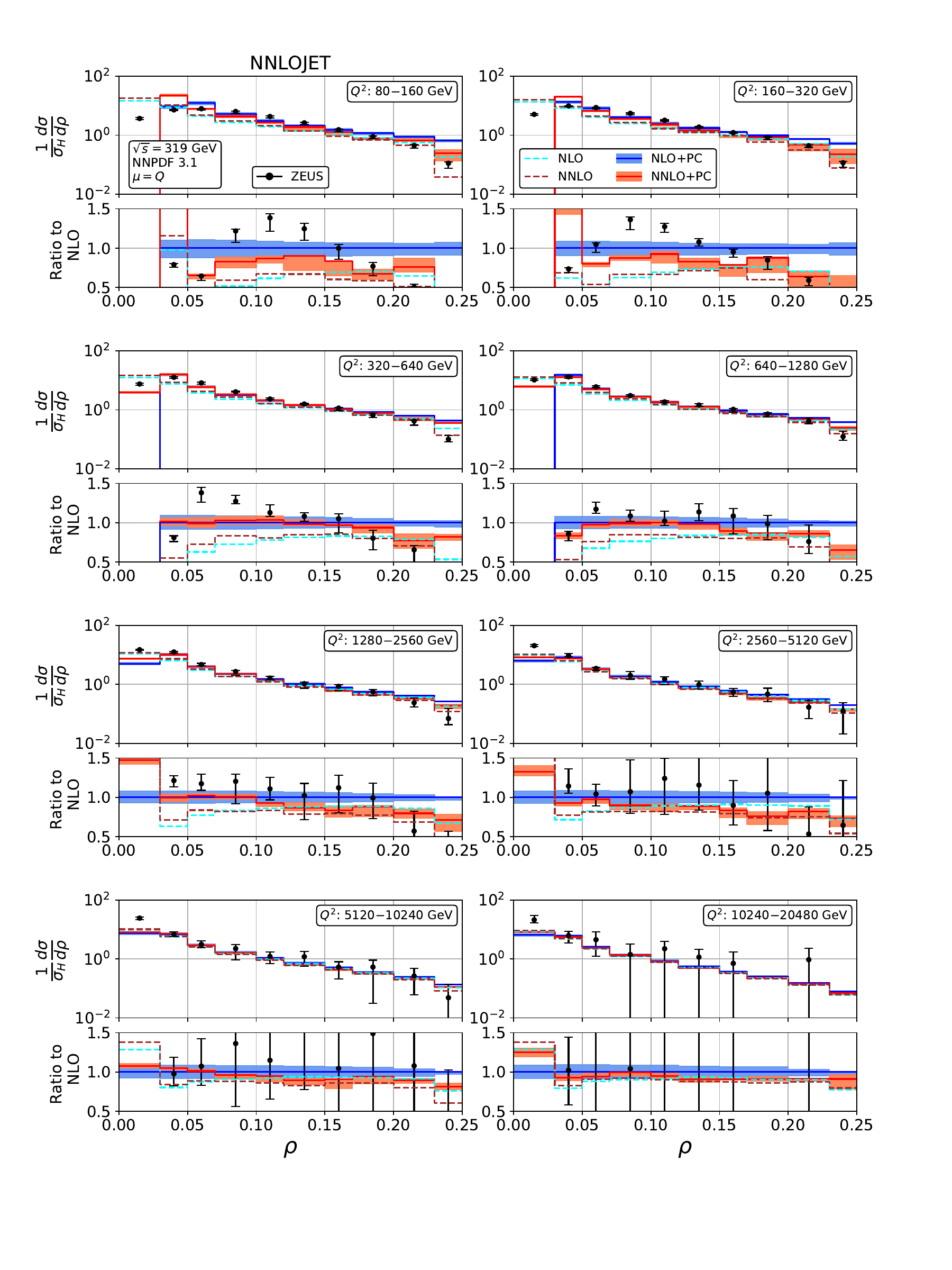}
  \caption{Event shape distribution for jet mass: $\rho$ fixed-order predictions at NLO (dashed cyan) and NNLO (dashed brown), and corrected 
  	for hadronization effects at NLO (blue) and NNLO (red), compared to ZEUS data~\protect{\cite{Chekanov:2006hv}}. The lower 
  frames display the ratio to the NLO prediction for the central scale $\mu^2=Q^2$.} 
  \label{fig:ZEUSTMP}
\end{figure}
\begin{figure}[t]
\hspace{-0.2in}   \includegraphics[width=3.8in,page=5]{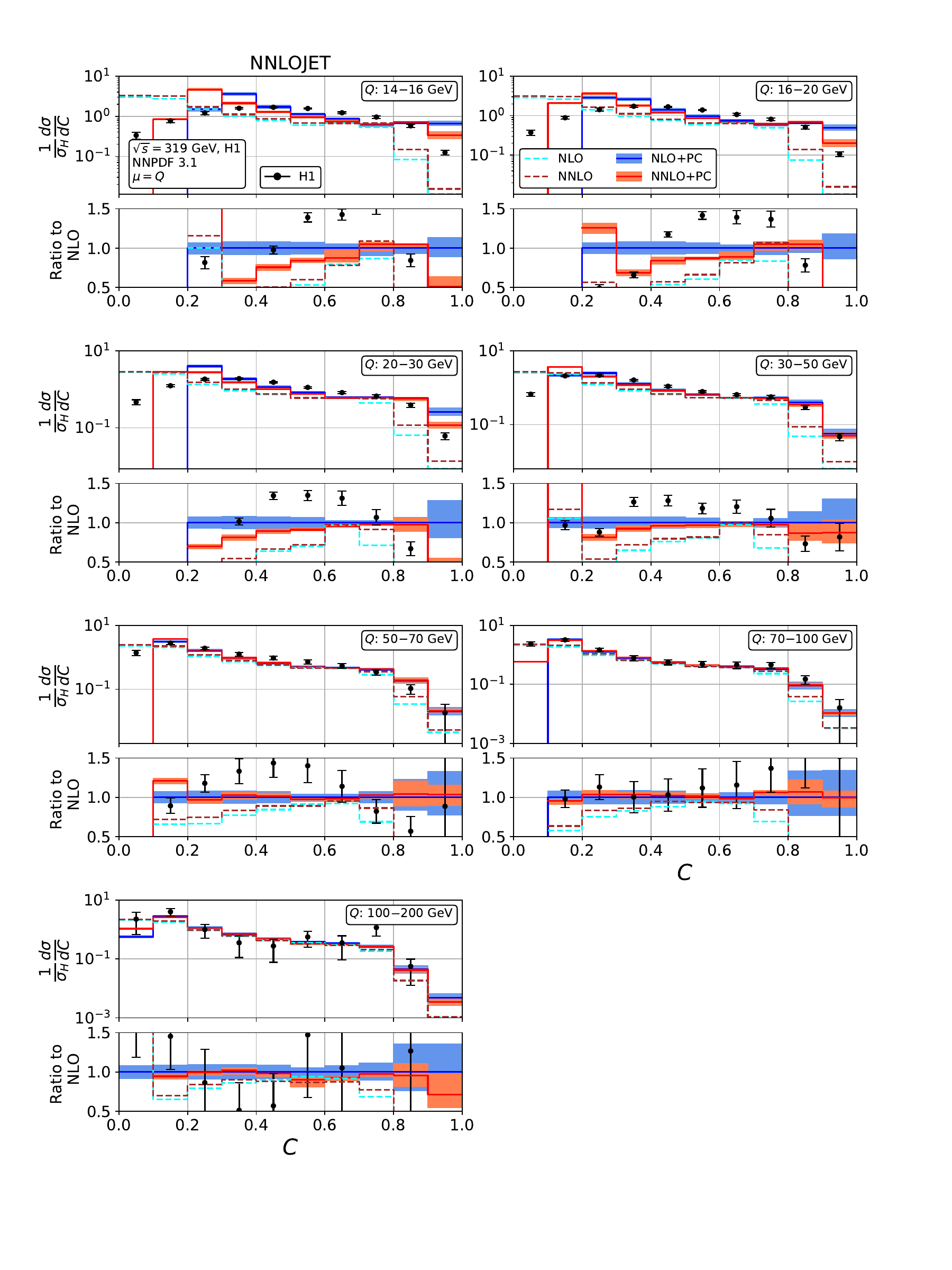}
  \caption{Event shape distribution for $C$-parameter: $C$  fixed-order predictions at NLO (dashed cyan), NNLO (dashed brown), and corrected 
  	for hadronization effects at NLO (blue) and NNLO (red), compared to H1 data~\protect{\cite{Aktas:2005tz}}. The lower 
  frames display the ratio to the NLO prediction for the central scale $\mu^2=Q^2$.} 
  \label{fig:H1TCP}
\end{figure}
\begin{figure}[t]
 \hspace{-0.2in}  \includegraphics[width=3.8in,page=6]{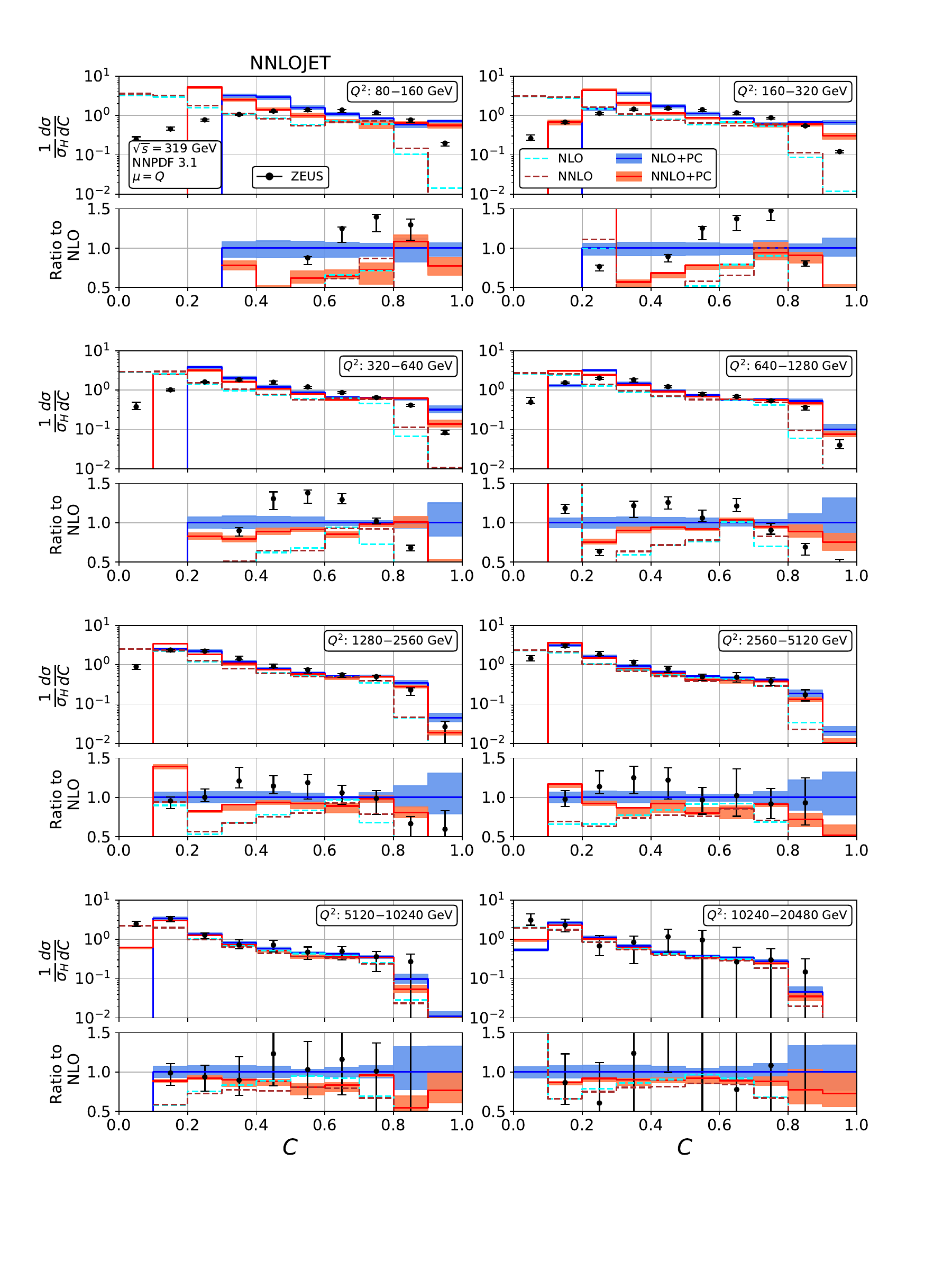}
  \caption{Event shape distribution for $C$-parameter: $C$ fixed-order predictions at NLO (dashed cyan) and NNLO (dashed brown), and corrected 
  	for hadronization effects at NLO (blue) and NNLO (red), compared to ZEUS data~\protect{\cite{Chekanov:2006hv}}. The lower 
  frames display the ratio to the NLO prediction for the central scale $\mu^2=Q^2$.} 
  \label{fig:ZEUSTCP}
\end{figure}
\begin{figure}[t]
\hspace{-0.2in}   \includegraphics[width=3.8in,page=2]{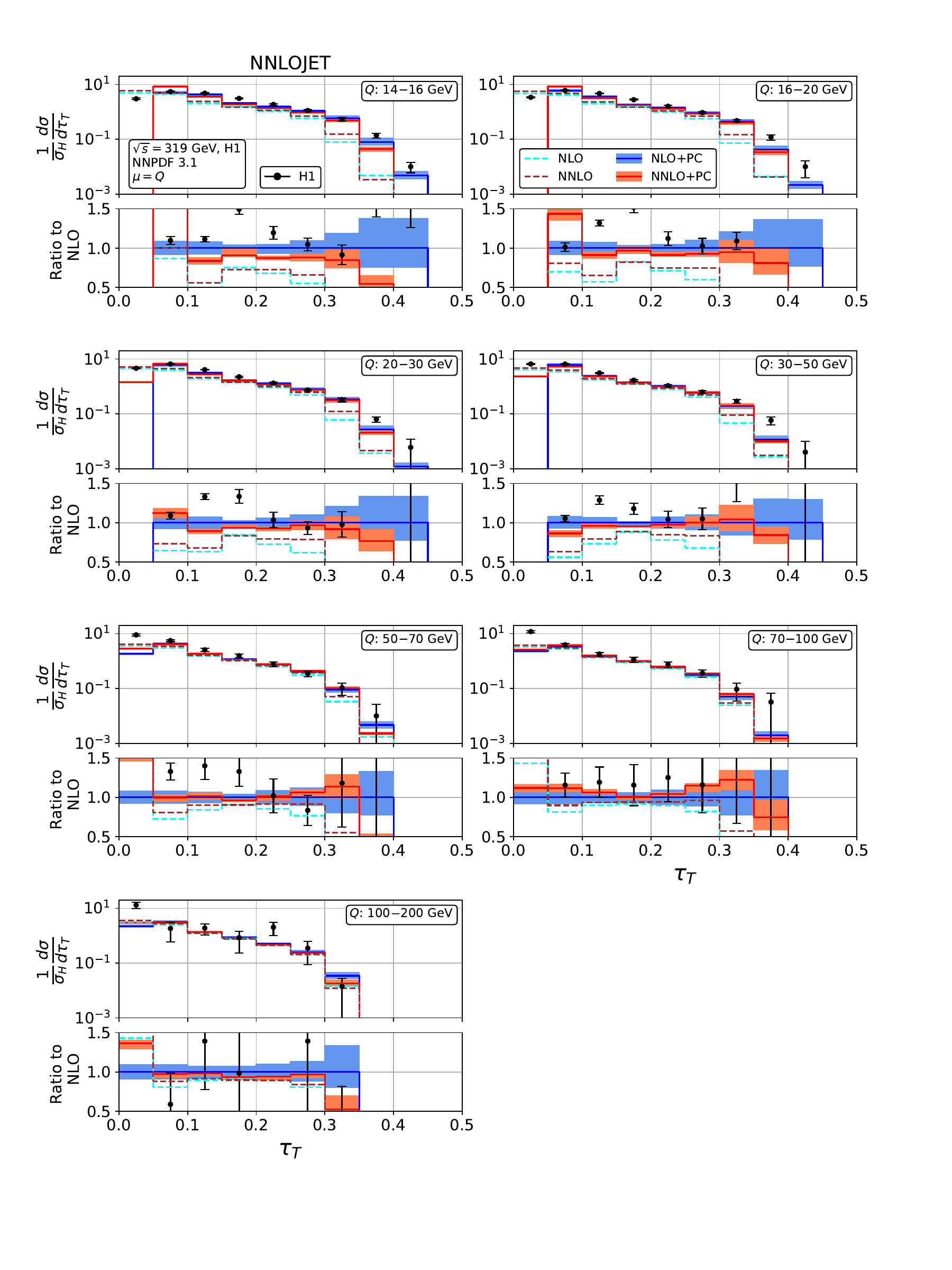}
  \caption{Event shape distribution for thrust with respect to thrust axis: $\tau_T$ fixed-order predictions at NLO (dashed cyan), NNLO (dashed brown), and corrected 
  	for hadronization effects at NLO (blue) and NNLO (red), compared to H1 data~\protect{\cite{Aktas:2005tz}}. The lower 
  frames display the ratio to the NLO prediction for the central scale $\mu^2=Q^2$.}
  \label{fig:H1TTP}
\end{figure}
\begin{figure}[t]
 \hspace{-0.2in}  \includegraphics[width=3.8in,page=2]{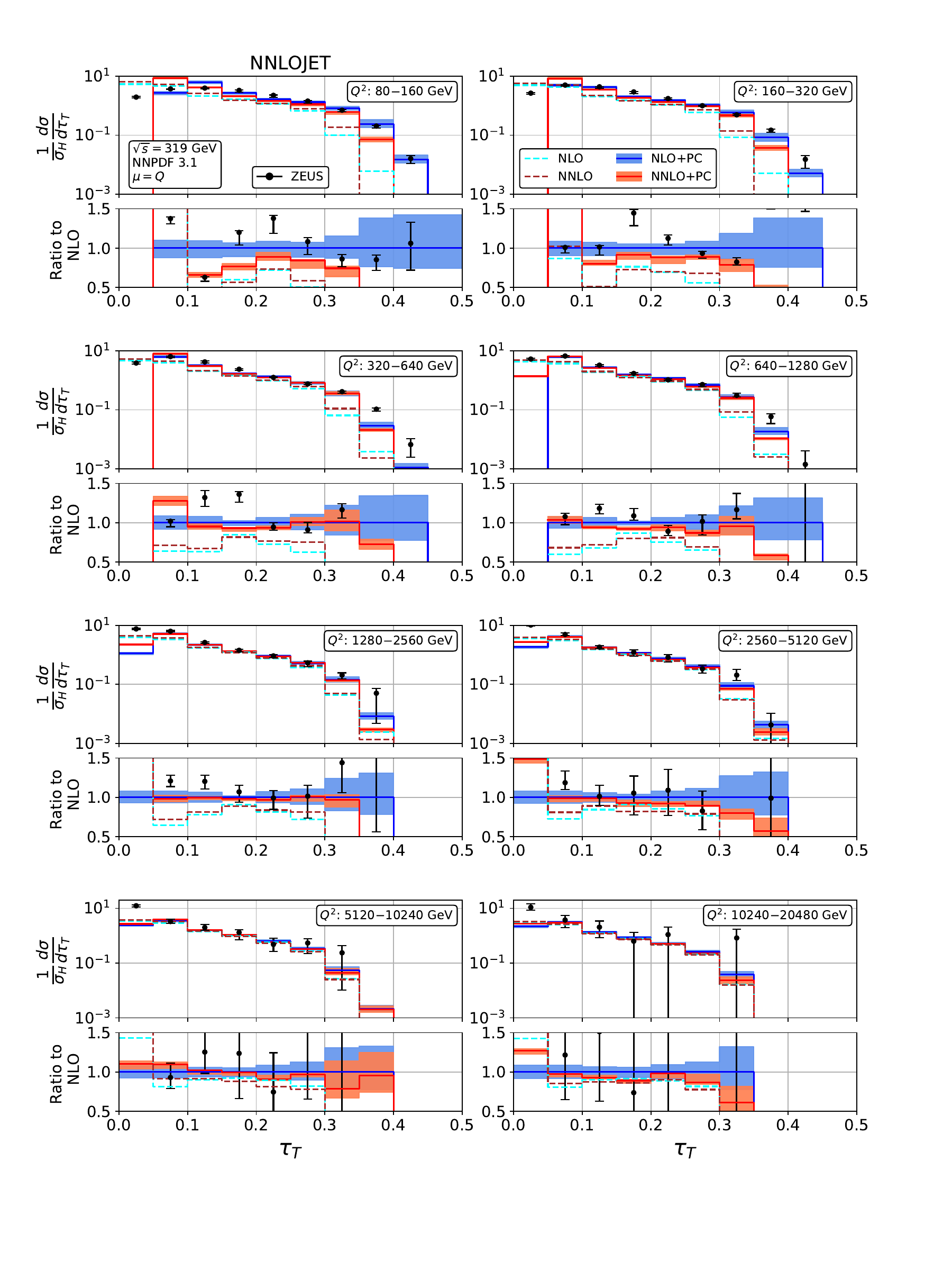}
  \caption{Event shape distribution for thrust with respect to thrust axis: $\tau_T$ fixed-order predictions at NLO (dashed cyan) and NNLO (dashed brown), and corrected 
  	for hadronization effects at NLO (blue) and NNLO (red), compared to ZEUS data~\protect{\cite{Chekanov:2006hv}}. The lower 
  frames display the ratio to the NLO prediction for the central scale $\mu^2=Q^2$.}
  \label{fig:ZEUSTTP}
\end{figure}
\begin{figure}[t]
\hspace{-0.2in}   \includegraphics[width=3.8in,page=4]{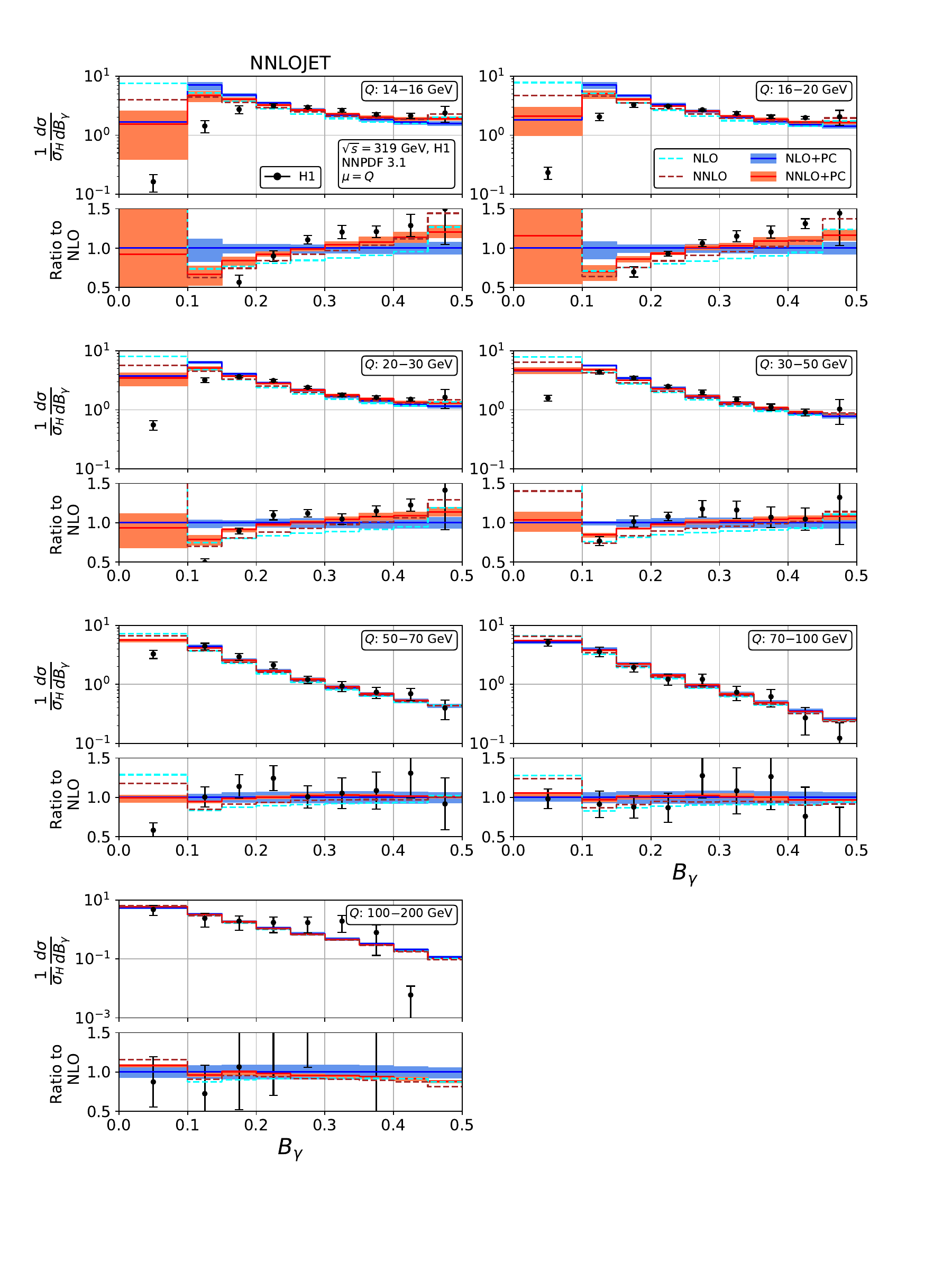}
  \caption{Event shape distribution for jet broadening   with respect to thrust axis: $B_\gamma$ fixed-order predictions at NLO (dashed cyan), NNLO (dashed brown), and corrected 
  	for hadronization effects at NLO (blue) and NNLO (red), compared to H1 data~\protect{\cite{Aktas:2005tz}}. The lower 
  frames display the ratio to the NLO prediction for the central scale $\mu^2=Q^2$.}
  \label{fig:H1BTP}
\end{figure}
\begin{figure}[t]
 \hspace{-0.2in}  \includegraphics[width=3.8in,page=4]{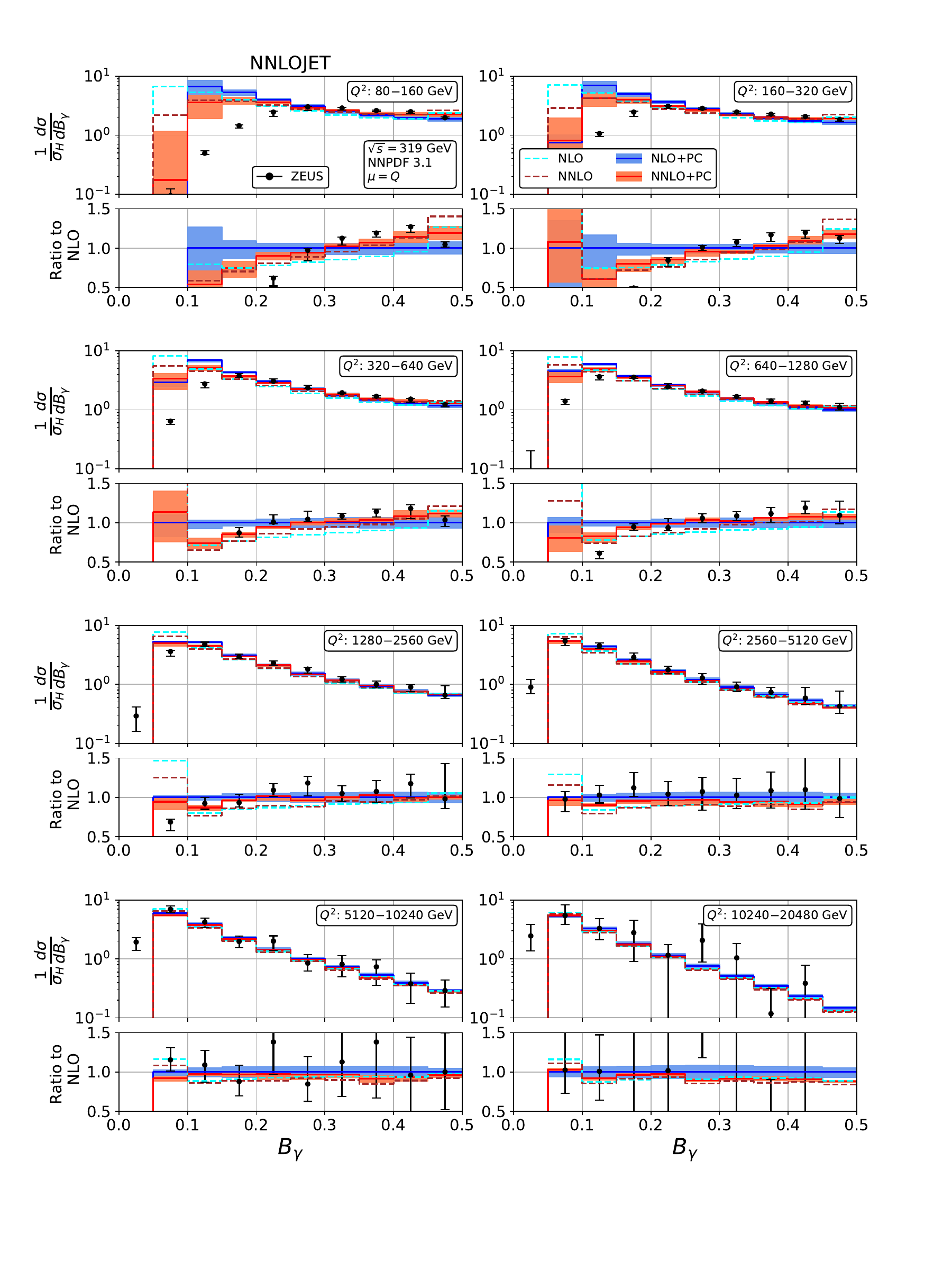}
  \caption{Event shape distribution for jet broadening   with respect to thrust axis: $B_\gamma$ fixed-order predictions at NLO (dashed cyan) and NNLO (dashed brown), and corrected 
  	for hadronization effects at NLO (blue) and NNLO (red), compared to ZEUS data~\protect{\cite{Chekanov:2006hv}}. The lower 
  frames display the ratio to the NLO prediction for the central scale $\mu^2=Q^2$.}
  \label{fig:ZEUSBTP}
\end{figure}

Figures~\ref{fig:H1TBP}--\ref{fig:ZEUSBTP} display the 
 theory predictions obtained by combining the fixed-order predictions up to NNLO with power corrections 
 computed  using the dispersive model as described in the previous section to the experimental data from H1~\cite{Aktas:2005tz} and ZEUS~\cite{Chekanov:2006hv}. 
 To illustrate the magnitude of the power corrections, the uncorrected fixed-order predictions for central scales $\mu=Q$ are indicated by blue lines at NLO and brown lines at NNLO.
 The shift~(\ref{eq:powershift}) is 
 applied on the high-resolution histograms (\ref{eq:hires}) which were computed with a lower cut-off affecting their first bin (where all-order
  resummation of large logarithmic corrections~\cite{Antonelli:1999kx,Dasgupta:2001eq,Dasgupta:2002dc,Dasgupta:2003iq} is  required to obtain 
 a finite prediction). The shifted high-resolution histograms are then combined to 
 the number of bins used in the experimental measurements:
 \ba
\tau_\gamma:& [0,1], &10\,, \nn\\
\tau_T:& [0,0.5], &10\,,\nn\\
\rho:& [0,0.25], &8(\mbox{H1})/10(\mbox{ZEUS})\,, \nn\\
B_\gamma:& [0,0.5], &10\,, \nn\\
C:& [0,1], &10\,.
\label{eq:expres}
\ea

Owing to the interplay of the lower cut-off on the distributions and the power correction shift, the prediction for the left-most non-vanishing 
bin of each distribution is unreliable, and should not be taken into account when 
comparing the experimental data with the theory predictions. A prediction for $F\to 0$ will have to include resummation in order to become meaningful. This limitation should be kept in mind in the following
comparisons. 

For the thrust distribution $\tau_\gamma$, Figures~\ref{fig:H1TBP} and 
\ref{fig:ZEUSTBP}, we observe that the NNLO corrections at low and moderate values of 
$Q^2$ are leading to an increase of the distribution in the bulk, and a slight decrease at high and low $\tau_\gamma$. 
At the highest values of $Q^2$, the NNLO corrections become very small and negative even in the bulk. Overall, the NNLO corrections improve the description of the data. 
Compared to NLO, inclusion of the NNLO correction leads to a reduction of the scale uncertainty. 
This reduction
is only moderate at the lowest values of $Q^2$, where the NNLO scale uncertainty remains at the 6\% level. At higher $Q^2$, the reduction of scale uncertainty at NNLO  is  
more pronounced, leading to predictions with residual uncertainty below 4\%. These uncertainties should be compared to the experimental errors. The ZEUS data~\cite{Chekanov:2006hv}
 are slightly more precise than the 
H1 data~\cite{Aktas:2005tz}, and also reach to lower values of $Q^2$.  In the low-$Q^2$ bins, the NNLO scale uncertainty remains larger than the experimental errors, as was also 
observed~\cite{Andreev:2016tgi} for 
jet production in DIS at low $Q^2$. For moderate and high values of $Q^2$, the scale uncertainty is now well below the experimental errors, thereby allowing for the 
use of the event shape distributions in precision QCD studies.

A similar pattern is also observed in the jet mass distribution, Figures~\ref{fig:H1TMP} and \ref{fig:ZEUSTMP}: positive NNLO corrections in the bulk at moderate $Q^2$, which turn
negative when going to large values of $\rho$ or to large $Q^2$. At the lowest values of $Q^2$, the NNLO corrections are negative throughout the distribution (despite positive corrections at parton-level) due 
to the reduced size of the power correction at NNLO. The NNLO corrections lead to an improved description of the shape of the experimental data. This improvement 
is particularly visible at large $\rho$ for all values of $Q^2$, where negative NNLO contributions lead to a considerably better description of the kinematical shape of the data.  
Overall, the 
agreement between data and theory is however somewhat worse for $\rho$ than it was for $\tau_\gamma$. The NNLO scale uncertainties also follow a similar pattern as for $\tau_\gamma$: compared to NLO
only a modest reduction at low $Q^2$ and a substantial reduction to the level of a few per cent at high $Q^2$. Again, the experimental errors are larger than the scale uncertainty for moderate and high $Q^2$, thus enabling precision QCD studies. 

In the $C$-parameter, Figures~\ref{fig:H1TCP} and 
\ref{fig:ZEUSTCP}, we must  distinguish the region below and above the Sudakov shoulder, which is located at $C=0.75$ in the perturbative parton-level 
expression. The dispersive power corrections shift the location 
of this shoulder to higher values of $C$. This shift is largest at low $Q^2$, and decreases in magnitude towards higher $Q^2$. The region above the Sudakov shoulder is kinematically forbidden at LO, and receives contributions only from NLO onwards. Already for values of $C$ below the Sudakov shoulder, the pattern of NNLO corrections is more intricate than 
what was observed in $\tau_\gamma$ and $\rho$.  The observed structure is due to 
presence of a kinematical ridge~\cite{Dasgupta:2002dc} in the perturbative expressions at $C\approx 0.515$, which destabilizes the 
perturbative convergence of the distribution in its vicinity, clearly visible in the high-resolution $C$-parameter distribution, Figure~\ref{fig:H1FO}. The perturbative predictions for the $C$-parameter 
distributions become quite precise at NNLO, with scale uncertainties of typically less than 8\% below the Sudakov shoulder and away from the kinematical ridge. They are however affected by large 
hadronization corrections, which shift the whole distribution by more than two bins in $C$ at low $Q^2$. 
Compared to all other event shape distributions, these power corrections are particularly large in the $C$-parameter distributions, see~(\ref{eq:apower}). For the lower values of $Q^2$, we also 
observe that the 
shape of the distribution is poorly described. For medium and large values of $Q^2$, 
the power corrections are much smaller, NNLO corrections are relatively small and uniform, and a satisfactory description of the experimental data is observed. 

The thrust distribution with respect to the thrust axis  $\tau_T$, Figures~\ref{fig:H1TTP} and 
\ref{fig:ZEUSTTP}, displays a similar pattern, with non-trivial structures in its perturbative expressions around the Sudakov shoulder at  $\tau_T=0.293$ and the kinematical ridge around $\tau_T\approx 0.13$,
which are both nicely visible in the $\tau_T$-distribution at high resolution, Figure~\ref{fig:H1FO}. These exceptional points are shifted to larger values of  $\tau_T$ by the power corrections. 
The NNLO corrections are negative and small throughout almost the whole distribution for all values of 
$Q^2$, except in the bin with the highest  $\tau_T$, which is already well above the Sudakov shoulder, and where the cross section is already very small. 
As for $\tau_\gamma$, we note that the smallness of the NNLO effect is mainly due to a cancellation between positive parton-level corrections and a decrease in the power corrections. 
The corrections are  typically within the 
NLO scale uncertainty band. 
The overall agreement between experimental data and theory predictions is satisfactory only at higher $Q^2$. Substantial discrepancies 
at low $Q^2$ are observed especially in the 
vicinity of the kinematical ridge, where the theory predictions are systematically and considerably below the data. This behaviour may be indicative of the need of an re-consideration of the perturbative 
expansion and of the hadronization corrections in regions around the kinematical ridges.

Finally, for the jet broadening with respect to the boson axis $B_\gamma$, Figures~\ref{fig:H1BTP} and 
\ref{fig:ZEUSBTP}, the NNLO corrections assume a non-trivial shape, changing from negative at small $B_\gamma$ to 
positive at large $B_\gamma$, thereby leading to a considerably better  description of the data. In these distributions, the onset of large logarithmic terms at low $B_\gamma$ is well visible, indicating the 
need for their resummation. The NNLO corrections lead to a considerable reduction of the scale uncertainty in the bulk of the distributions, which is more pronounced at low $Q^2$ than in most other 
event shape distributions. The NNLO scale uncertainty ranges from 8\% at low $Q^2$ to 3\% at high $Q^2$, which is comparable to or below the experimental uncertainties throughout.

Across the different event shape distributions,  several common  features are observed. The NNLO corrections are typically moderate, and fall usually within the NLO scale uncertainty bands. This is particularly remarkable since the NLO corrections were typically large (often comparable in size to the LO predictions), and well outside the LO scale uncertainty bands. The numerical smallness of the NNLO effect is 
often due to a partial cancellation between the parton-level correction and modification of the power corrections at this order. 
Except in the low-$F$ region, where 
large logarithmic corrections require an all-order resummation, and in the vicinity of Sudakov shoulders and kinematical ridges, we observe the onset of a good convergence of the perturbative fixed-order expansion. The corrections at low values of $Q^2$ are inevitably larger (due to the larger expansion parameter), which also translates in a sizeable scale uncertainty remaining at NNLO of about 10\%. At 
larger values of $Q^2$, this scale uncertainty improves considerably to the typically 4\% or below, clearly highlighting the potential of precision QCD studies with event shapes based on 
existing HERA data~\cite{Aktas:2005tz,Chekanov:2006hv}, or with much larger data sets for hadronic final states that could be obtained 
at a future electron-ion collider~\cite{Accardi:2012qut} or at the LHeC~\cite{AbelleiraFernandez:2012cc}. The non-perturbative power corrections that we obtained in the dispersive model induce 
large shifts in some of the distributions, especially at low $Q^2$, where the statistical quality of the data is largest. Moreover, their application to  the distributions in the form of a constant shift 
is only an approximation, which should be revisited carefully as soon as more precise data are becoming available.

\subsection{Mean values}
\begin{figure}[t]
	\centering
	\includegraphics[width=3.5in]{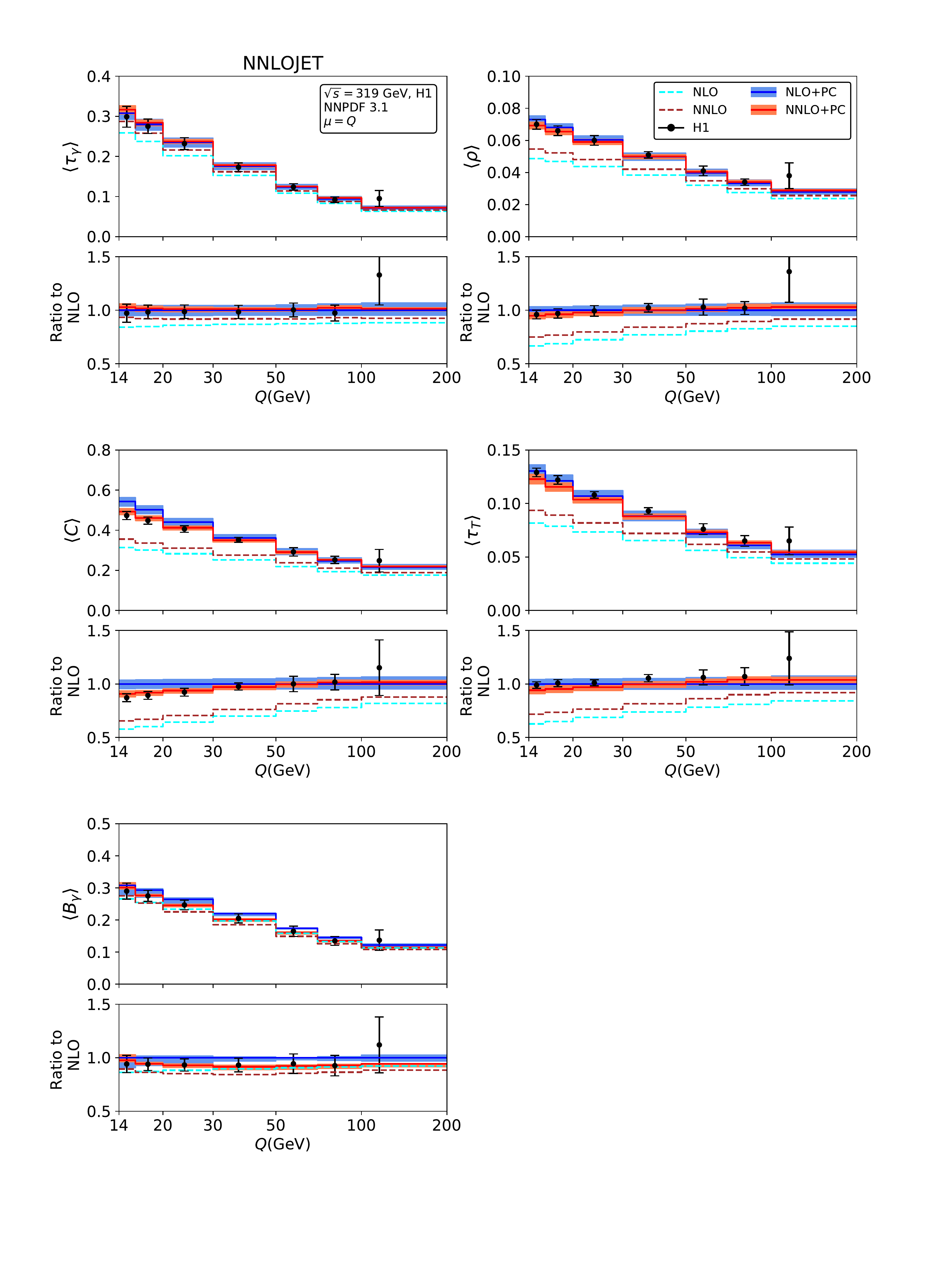}
	\caption{Mean value of the event shapes at NLO (blue) and NNLO (red) including power corrections, compared to H1 data~\protect\cite{Aktas:2005tz}. The fixed-order NLO and NNLO predictions 
	(dashed cyan and brown lines) are included to illustrate the magnitude of the power corrections.}
	\label{fig:H1meanFOPC}
\end{figure}

\begin{figure}[t]
	\centering
	\includegraphics[width=3.5in]{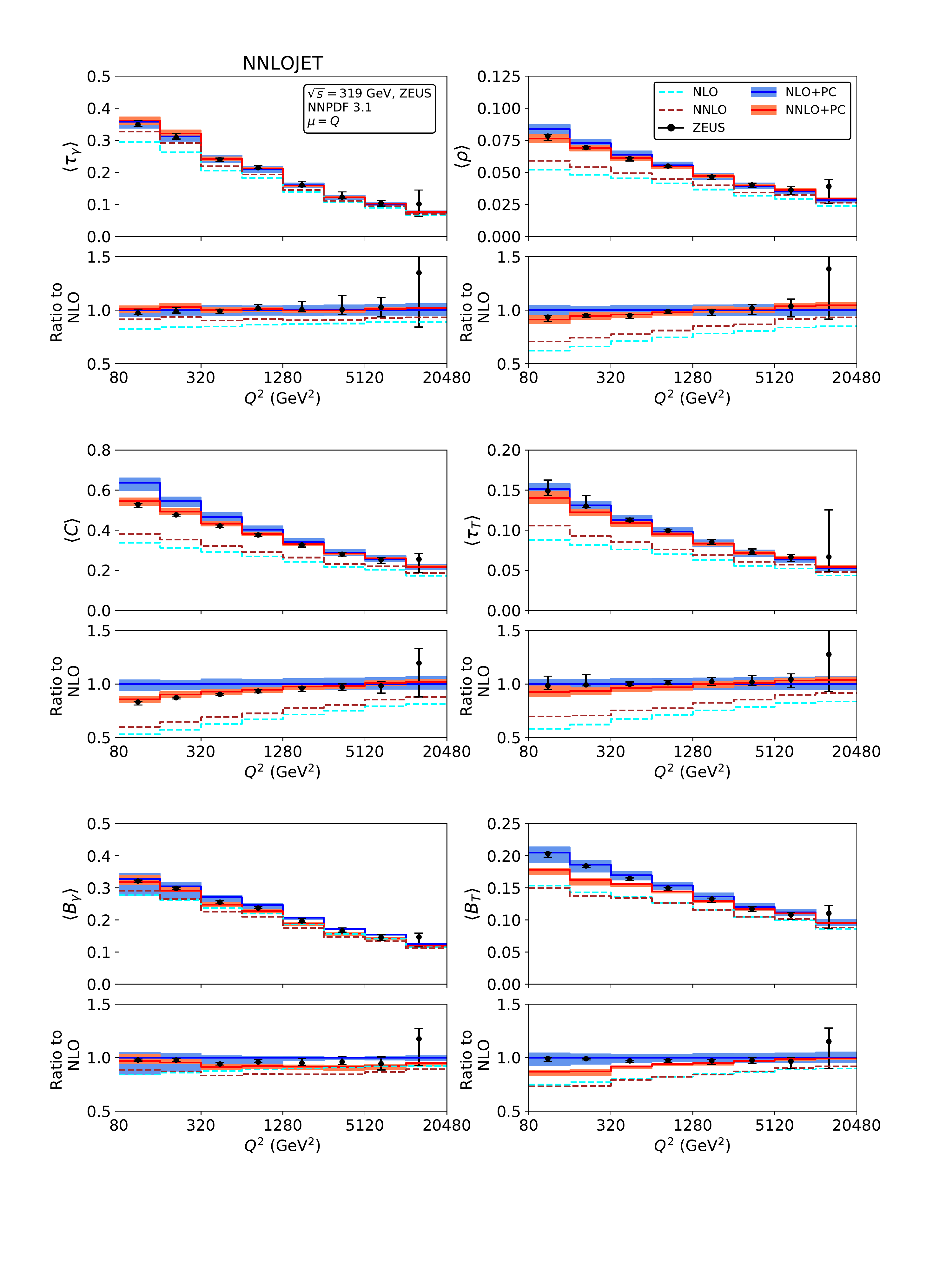}
	\caption{Mean value of the event shapes at NLO (blue) and NNLO (red) including power corrections, compared to ZEUS data~\protect\cite{Chekanov:2006hv}. The fixed-order NLO and NNLO predictions 
	(dashed cyan and brown lines) are included to illustrate the magnitude of the power corrections.} 
		\label{fig:ZEUSmeanFOPC}
\end{figure}

The power corrections to the mean values result in an additive shift  of the perturbative predictions, see (\ref{eq:powermean}). As for the event shape distributions, we truncate $P$ in (\ref{eq:powerP})
to order $\alpha_s^2(Q)$ 
for the power corrections to the NLO fixed order predictions and to $\alpha_s^3(Q)$ for power corrections applied to the NNLO predictions. 
Applying this shift to the perturbative results of Section~\ref{sec:resultmean}, 
we obtain hadron-level predictions for the mean values, which are compared to the data from H1~\cite{Aktas:2005tz} and ZEUS~\cite{Chekanov:2006hv} in Figures~\ref{fig:H1meanFOPC} and~\ref{fig:ZEUSmeanFOPC}. The fixed-order predictions for central scales $\mu=Q$ are indicated by blue lines at NLO and brown lines at NNLO, showing that the power corrections are sizeable  
 for all mean values. Their inclusion eliminates the tension between data and purely perturbative results seen in 
Figures~\ref{fig:H1meanFO} and \ref{fig:ZEUSmeanFO} above. 

Comparing the mean values with and without power corrections, we observe 
that the large positive NNLO corrections at low $Q^2$ that are seen in Figures~\ref{fig:H1meanFO} and \ref{fig:ZEUSmeanFO} are more than compensated by the decrease in the 
numerical magnitude of the power correction in going from NLO to NNLO in $P$. The combined effect of the NNLO contributions to the fixed order predictions and the power corrections is 
typically a small reduction of the mean values at low $Q^2$, thereby leading to an improved description of the H1 and ZEUS data. At larger $Q^2$, this combined effect results in a very small 
change of the predictions from NLO to NNLO, which comes with a substantial reduction of the perturbative scale uncertainty, which is almost halved. 

Both the  H1~\cite{Aktas:2005tz} and ZEUS~\cite{Chekanov:2006hv} studies used their measurements of event shape distributions for a simultaneous fit of
the QCD coupling constant $\alpha_s(M_Z)$ 
and the effective coupling $\alpha_0$ that appears in the power correction, performed using NLO fixed-order results. While ZEUS lists only the results obtained 
for the individual 
event shape variables (displaying a substantial scatter), H1 also performed a combined fit, resulting in 
 $\alpha_s(M_Z) = 0.1198\pm 0.0012 {\rm (exp)} {+0.0056 \atop -0.0043} {\rm (th)}$ and 
$\alpha_0 = 0.476\pm0.008  {\rm (exp)} {+0.018 \atop -0.059} {\rm (th)}$. Especially the theory error on $\alpha_s(M_Z)$ is largely dominated by the NLO scale uncertainty. 

A similar NNLO study has been performed previously on event shape moments in $e^+e^-$ annihilation~\cite{Gehrmann:2009eh}, resulting in decreased scatter between 
different shape variables, and in lower theory uncertainties. Also, a slight increase in the fitted central value of $\alpha_0$ from NLO to NNLO was observed in $e^+e^-$, resulting in an NNLO value 
of $\alpha_0 =0.5132\pm 0.0115\mbox{(exp)}\pm 0.0381\mbox{(th)}$. 
Our predictions in Figures~\ref{fig:H1meanFOPC} and \ref{fig:ZEUSmeanFOPC} use $\alpha_0 =0.5$ throughout. They result in a very good description of the data at NNLO, and we 
notice that the NLO curves could be brought into better agreement with the 
data by a slight lowering of $\alpha_0$, towards its H1 fit value. 

With the newly derived NNLO corrections, the combined fit of $\alpha_s(M_Z)$  and $\alpha_0$  to  event shape distributions and their mean values can now be repeated fully consistently to NNLO accuracy. 
It can be anticipated that the 
main effect of the NNLO corrections will be in a reduction of the theory-induced uncertainty on the extracted value of $\alpha_s(M_Z)$, which was found to be about 5\% in the NLO-based study by 
H1~\cite{Aktas:2005tz}. This reanalysis of the experimental data will require our fixed-order results to be re-cast into 
convolution grids~\cite{Britzger:2019kkb} that enable an efficient re-evaluation for multiple parameter values and parton distributions, which  is  beyond the scope of the present paper.

\section{Conclusions}
\label{sec:conc}

In this paper, we computed the NNLO QCD corrections to event shape distributions and their mean values in deep inelastic lepton-proton scattering. Our calculation was performed 
in the \NNLOJET framework, and is largely based on the NNLO corrections~\cite{Currie:2016ytq,Currie:2017tpe} to di-jet production in DIS. The NNLO corrections to the distributions are not uniform, 
although some general trends are observed: positive corrections in the bulk of the distributions at low and medium $Q^2$, negative corrections in the bulk at high $Q^2$ and 
 at the upper kinematical boundaries of the shape variables for all $Q^2$. Several perturbative instabilities due to Sudakov shoulders~\cite{Catani:1997xc} or kinematical ridges~\cite{Dasgupta:2002dc}
were observed in $C$ and $\tau_T$. 
Predictions in the kinematic vicinity of these exceptional points will require novel resummation approaches to overcome the associated  instabilities of the fixed-order predictions.
Moreover, at low values of the event shape variables, we observed the onset of large logarithmic corrections at each order in perturbation theory. These were particularly pronounced in 
$B_\gamma$. Resummation of these corrections is currently understood to next-to-leading logarithmic accuracy~\cite{Dasgupta:2003iq}. Aiming for a matching 
between fixed order and resummation in a form where the fixed-order expansion of the resummation formula reproduces all logarithmically enhanced terms up to NNLO 
(as was done for $e^+e^-$ event shapes~\cite{Becher:2008cf,Abbate:2010xh,Hoang:2014wka}) will require two more logarithmic 
orders in the resummation. 

To compare our parton-level predictions with hadron-level data, we used the dispersive model~\cite{Dokshitzer:1995qm} to estimate the leading power correction effects from hadronization. 
The model is based on
the study two-point correlators which relate to the mean values of the event shape distributions.  
 On the 
event shape distributions, additional assumptions must be made concerning the kinematical dependence of the power corrections. The power correction factors receive higher order contributions in 
the strong coupling constant, which we truncate to the same level as used in the fixed-order parton-level predictions. 

Our resulting hadron-level predictions were compared to data from the H1~\cite{Aktas:2005tz} and ZEUS~\cite{Chekanov:2006hv} experiments. On  the event shape distributions, we 
observe that inclusion of the NNLO corrections leads in general to an improved description of their kinematical shapes. Especially at medium and high $Q^2$, the NNLO corrections result in 
a substantial reduction of the scale uncertainties of the predictions, to the level of a few per cent. A similar reduction of scale uncertainty is also observed on the mean values.
On these mean values, we observe a compensation between the positive NNLO corrections to the fixed-order parton-level predictions and the negative NNLO contributions to the power 
corrections, resulting in a relatively small net effect at NNLO. 
Our newly derived NNLO results yield predictions with scale uncertainties that are typically below the experimental errors of the available HERA data on event shape distributions. 
They motivate a full NNLO-based reanalysis of event 
shape distributions and mean values. This should be leading to an improved determination of $\alpha_s(M_Z)$ and $\alpha_0$, 
which was previously limited by the uncertainty on the NLO theory.  

With high-resolution measurements 
of event shape distributions in deep inelastic scattering at a future electron-ion collider~\cite{Accardi:2012qut} or at the LHeC~\cite{AbelleiraFernandez:2012cc}, our results will
enable a broad spectrum of precision QCD studies, aiming for an improved understanding of its perturbative and non-perturbative aspects. 

\section*{Acknowledgements}
The authors thank Xuan Chen, Juan Cruz-Martinez, James Currie, Rhorry Gauld, Aude Gehrmann-De Ridder, Nigel Glover, Marius H\"ofer, Imre Majer, 
Tom Morgan, Joao Pires and James Whitehead for useful discussions and their many contributions to the \textsc{NNLOjet} code. 
This research was supported in part by the UK Science and Technology Facilities Council under contract ST/G000905/1, by the Swiss National Science Foundation (SNF) under contract 200020-175595, by the ERC Consolidator Grant HICCUP (614577) and by the  ERC Advanced Grant MC@NNLO (340983).

\bibliography{DIS_shapes}

\end{document}